\documentclass[12pt,letterpaper]{article}
\pdfoutput=1
\usepackage{soul}

\usepackage{jheppub}
\usepackage{amsfonts}
\usepackage{amsmath}
\allowdisplaybreaks[4]         
\usepackage{amssymb}   
\usepackage{euscript}       
\usepackage[dvipsnames]{xcolor}
\usepackage{IEEEtrantools}
\usepackage{upgreek}
\usepackage{graphicx}
\usepackage{caption}
\usepackage{subcaption}
\usepackage[export]{adjustbox}

\usepackage{tensor}
\usepackage{braket}
\usepackage{float}

\newcommand{\EE}{\mathrm{EE}}

\usepackage{hyperref}
\hypersetup{
	colorlinks =WildStrawberry,
	linkcolor =red,
	pdftitle={Shape dependence of renormalized entanglement entropy},
	citecolor=MidnightBlue,
	filecolor=WildStrawberry,
	urlcolor=WildStrawberry,
}

\newcommand{\bea}{\begin{eqnarray}}
\newcommand{\eea}{\end{eqnarray}}
\newcommand{\ba}{\begin{eqnarray}}
\newcommand{\ea}{\end{eqnarray}}

\newcommand{\beq}{\begin{equation}}
\newcommand{\eeq}{\end{equation}}
\newcommand{\beqa}{\begin{eqnarray}}
\newcommand{\eeqa}{\end{eqnarray}}
\newcommand{\beqar}{\begin{eqnarray*}}
\newcommand{\eeqar}{\end{eqnarray*}}

\newcommand{\e}[1]{\text{e}^{#1}}

\newcommand{\diff}{\mathrm{d}}
\newcommand{\vol}{\text{Vol}}
\newcommand{\ren}{\text{ren}}
\newcommand{\rt}{\text{RT}}

\renewcommand{\href}[2]{#2}

\arxivnumber{20nn.nnnnn}
\title{Shape dependence of renormalized holographic entanglement entropy\\
}
\author[a]{Giorgos Anastasiou,}
\author[a]{Javier Moreno,} 
\author[b]{Rodrigo Olea}
\author[c]{and David Rivera-Betancour}

\affiliation[a]{Instituto de F\'isica, Pontificia Universidad Cat\'olica de Valpara\'iso,\\ Casilla 4059, Valpara\'iso, Chile.}
\affiliation[b]{Departamento de Ciencias F\'isicas, Universidad Andres Bello,\\
Sazi\'e 2212, Piso 7, Santiago, Chile}
\affiliation[c]{Centre de Physique Th\'eorique, CNRS, \'Ecole Polytechnique,\\*91128 Palaiseau, Cedex, France.}

\vspace{0.1cm}
\emailAdd{georgios.anastasiou@pucv.cl, francisco.moreno.g@mail.pucv.cl, rodrigo.olea@unab.cl, david.rivera-betancour@zimbra.polytechnique.fr}

\abstract{We study the holographic entanglement entropy of deformed entangling regions in three-dimensional CFTs dual to Einstein-AdS gravity, using a renormalization scheme based on the addition of extrinsic counterterms. In this prescription, when even-dimensional manifolds are considered, the universal contribution to the entanglement entropy is identified as the renormalized volume of the Ryu-Takayanagi hypersurface, which is written as the sum of a topological and a curvature term. It is shown that the change in the renormalized entanglement entropy due to the deformation of the entangling surface is encoded purely in the curvature contribution. In turn, as the topological part is given by the Euler characteristic of the Ryu-Takayanagi surface, it remains shape independent. Exploiting the covariant character of the extrinsic counterterms, we apply the  renormalization scheme for the case of deformed entangling regions in AdS$_4$/CFT$_3$, recovering the results found in the literature. Finally, we provide a derivation of the relation between renormalized entanglement entropy and Willmore energy. The presence of a lower bound of the latter makes manifest the relation between the AdS curvature of the  Ryu-Takayanagi surface and the strong subadditivity property. 

}

\begin{document}

\maketitle

\newpage
\section{Introduction}

Entanglement Entropy (EE) has attracted great interest in recent literature, as it appears in areas of theoretical physics as diverse as quantum information, condensed matter and quantum gravity. It also unveils unexpected links between these fields (see refs.\citep{Ryu:2006ef, Amico:2007ag, Calabrese:2009qy, Casini:2009sr, Rangamani:2016dms,Nishioka:2018khk,Witten:2018lha} for reviews on the subject).

In the context of gauge/gravity duality, the Ryu-Takayanagi (RT) formula \citep{Ryu:2006bv} relates the EE of a entangling subregion in a Conformal Field Theory (CFT) with the area of a codimension-2 hypersurface immersed in Einstein-anti-de Sitter (AdS) spacetime. This relation was proven in ref.\citep{Lewkowycz:2013nqa}.

This idea has inspired extensive work in the subject, full of appealing relations and conjectures. Some concrete examples include the emergence of spacetime from the first law of entanglement entropy \citep{VanRaamsdonk:2009ar} and the proposed solution of the firewall paradox \citep{Maldacena:2013xja}. In the case of non trivial topologies, the entropy of de Sitter space was recently interpreted as the holographic entanglement entropy between two disconnected conformal boundaries \citep{Arias:2019pzy}.


In the CFT side, in ref.\citep{Calabrese:2009qy} it was shown that EE is obtained at the limit of R\'enyi entropy when the replica parameter $m$ tends to the unity. The introduction of the replica trick led Lewkowycz and Maldacena \citep{Lewkowycz:2013nqa,Dong:2016fnf} to consider a squashed-cone $(d+1)$-dimensional replica orbifold $\mathcal{M}_{d+1}^{(\alpha)}$. That is, in the bulk gravity side, a conically singular manifold without U(1) symmetry. Here, $\alpha$ is a conical angular variable such that the cone has an angular deficit given by $2\pi(1-\alpha)$, and related to the replica parameter by $\alpha=\frac{1}{m}$. Based on these considerations, the EE is defined as
\begin{equation}\label{REdef}
S_{\EE}=-\lim_{\alpha\rightarrow 1}\partial_\alpha I_{\text{E}}\left[\mathcal{M}_{d+1}^{(\alpha)}\right],
\end{equation}
where $I_{\text{E}}\left[\mathcal{M}_{d+1}^{(\alpha)}\right]$ is the Euclidean action evaluated on the orbifold $\mathcal{M}_{d+1}^{(\alpha)}$. By definition, R\'enyi entropy considers an integer replica parameter. The fact that it is related to the aperture of the cone allows for non-integer values, such that the limit \eqref{REdef} is well defined. In the particular case of $I_{\text{E}}$ being the Einstein-Hilbert (EH) action, the limit reproduces the RT formula for the EE.

Let $A$ be a smooth entangling region on a time slice of  a $d$-dimensional CFT, the general form of the EE is given by the expansion \cite{Grover:2011fa,Liu:2012eea}
\begin{equation}\label{Seven}
S_{\EE}(A)=c_{d-2}\frac{l^{d-2}}{\delta^{d-2}}+c_{d-4}\frac{l^{d-4}}{\delta^{d-4}}+\ldots+\begin{cases}
c_2\frac{l^2}{\delta^2}+s_{\text{univ}}(A) \log \frac{l}{\delta}+c_{0} & \text{for even }d, \\
c_1\frac{l}{\delta}+(-1)^{(d-1)/2}s_{\text{univ}}(A) & \text{for odd }d.
    \end{cases}       
\end{equation}
In this expression, $\{c_i\}$ are scheme-dependent coefficients. Thus, they are not physically observable. In turn, $\delta$ and $l$ are the energy cut-off of the theory and a characteristic scale of the entangling region, respectively.


In even dimensions, $s_{\text{univ}}$ is a linear combination of local integrals on the entangling surface, whose coefficients corresponds to the conformal anomaly of the theory (see refs.\cite{Calabrese:2004eu,Calabrese:2009qy,Solodukhin:2008dh} for examples). If the entangling surface is spherical, the only contribution to $s_{\text{univ}}$ comes from the type $A$-anomaly. On the other hand, if it is cylindrical, the surviving contributions come from the $B$-type ones \cite{Nishioka:2018khk}. For odd-dimensional CFTs, the lack of logarithmic term reflects the absence of conformal anomaly. Nevertheless, the finite part is physically relevant. It is also shown to be highly non-local, as opposed to the even-dimensional case. Interestingly, when computed for ball-shaped entangling regions, $s_{\text{univ}}$ is equivalent to the free energy $F_{\mathbb{S}^d}$ of a CFT placed on $\mathbb{S}^d$ background \citep{Casini:2011kv,Dowker:2010yj}. The sign is introduced in order to maintain positivity regardless the dimension of the CFT \citep{Klebanov:2011gs}.



In the particular case of CFT$_3$ on $\mathbb{S}^3$, the free energy is a monotonic function of the energy under Renormalization Group (RG) flows \citep{Jafferis:2011zi,Myers:2010xs,Myers:2010tj,Casini:2012ei}. For this reason, it is considered as an $F$-function\footnote{It has been proposed that the $F$-theorem holds also to higher dimensions, but no definite proof has been provided \citep{Giombi:2014xxa,Fei:2014yja,Jafferis:2012iv}.}, a measure of the number of degrees of freedom of the effective field theory at a certain energy scale \citep{Polchinski:1983gv,Wilson:1973jj}. This fact makes the universal term $s_{\text{univ}}$, evaluated at a circular entangling surface, a valuable probe of the $F$-theorem.

For arbitrary entangling regions, physical information of the field theory can be extracted from its shape. Studies on smooth entangling regions with symmetry can be found in refs.\cite{Allais:2014ata,Lewkowycz:2014jia,Fonda:2015nma}. In the case of non-smooth entangling regions, the expansion \eqref{Seven} is modified and new universal contributions to EE emerge \cite{Casini:2006hu,Hirata:2006jx,Klebanov:2012yf,Myers:2012vs,Kallin:2014oka,Bueno:2019mex}.

The shape dependence of EE is also studied perturbatively around maximally symmetric entangling regions in refs.\cite{Mezei:2014zla,Rosenhaus:2014woa,Bueno:2015lza,Bianchi:2015liz,Faulkner:2015csl,Bianchi:2016xvf,Dong:2016wcf,Ghosh:2017ygi,Carmi:2015dla,Jang:2020cbm}. More specifically, in refs.\citep{Allais:2014ata,Mezei:2014zla}, it is shown that the EE of a spherical entangling surface with deformations $\mathbb{S}^{1}_\epsilon$ in a CFT adopts the expansion
\begin{equation}\label{eq:introdef}
S_{\EE}^{\ren}(\mathbb{S}^{1}_\epsilon)=S^{\ren,(0)}_{\EE}(\mathbb{S}^{1})+\epsilon^2 S^{\ren,(2)}_{\EE}(\mathbb{S}^{1})+\mathcal{O}(\epsilon^3).
\end{equation}
Here, $\epsilon$ is a small deformation parameter and $S^{\ren,(0)}_{\EE}$ is the renormalized EE of the unperturbed sphere. The linear term in $\epsilon$ vanishes as the sphere is a minimum of the universal term amongst all shapes. The subleading term in expansion \eqref{eq:introdef} is proportional to the coefficient $C_T$ of the two-point function of the stress tensor
\begin{equation}\label{eq:propto}
S^{\ren,(2)}_{\EE}(\mathbb{S}^{1})\propto C_T,
\end{equation}
where
\begin{equation}
\braket{T_{ij}(x)T_{kl}(0)}=\frac{C_T}{x^{2d}}\left[I_{i(k}I_{l)j}-\frac{\delta_{ij}\delta_{kl}}{d}\right],
\end{equation}
and $I_{ij}=\delta_{ij}-2\frac{x_i x_j}{x^2}$. In the case of three-dimensional CFTs dual to Einstein gravity, the coefficient is given by $C_T=\frac{3L^2}{\pi^{3}G}$.

These holographic results for deformed entangling surfaces were extended to arbitrary dimensions in refs.\cite{Allais:2014ata,Mezei:2014zla}. They were later supported by field theory computations \citep{Faulkner:2015csl}.


The connection between the renormalized EE and renormalized volume of dual RT surface \citep{Anastasiou:2018rla}, provides a novel geometric interpretation on the origin of the shape-dependent terms. When a bulk AdS$_4$ spacetime is considered, the renormalized area of the RT surface is associated to the Willmore energy of a closed manifold immersed on $\mathbb{R}^3$ \cite{Babich:1992mc,Fonda:2015nma}. A similar connection between the Willmore energy and the renormalized volume have been provided earlier in mathematical literature \cite{alexakis2010renormalized}.

The Willmore energy is a geometrical quantity that measures the deviation of a closed surface from sphericity in $\mathbb{R}^3$ \cite{marques2014willmore,willmore1996riemannian,toda}. It has appeared in different fields of study, even beyond mathematics and physics. Applications of it can be found in biology, in order to study elastic properties of cell membranes (along with its generalization, the Helfrich energy \cite{helfrich1973elastic}). It also applies to computer graphics \cite{lott1988method} and mesh processing \cite{botsch2010polygon}. In the context of holographic EE, we propose it as a useful probe of the shape deformations of an entangling region. As it is a functional invariant under conformal transformations of the ambient metric, it induces a conformal structure. This will play an important role on our analysis.

Motivated by the results outlined above, in this paper, we study the shape dependence of the EE and its connection to Willmore energy. This paper is organized as follows:

In section \ref{Ren}, we review Kounterterms renormalization scheme in connection to the renormalized EE for a spherical entangling region developed in refs.\citep{Anastasiou:2017xjr,Anastasiou:2018mfk,Anastasiou:2018rla}. This quantity is found to be proportional to the renormalized volume of the RT surface. Following this idea, we provide additional examples and the corresponding interpretation of the results.


In section \ref{RenEx}, we focus on an entangling region which is a deformed disk  in CFT$_3$ to compute holographically the renormalized EE following the scheme described in section \ref{Ren}. We obtain a formula that reads
\begin{equation}
S_{\EE}^\ren \left(A\right)=-\frac{\pi L^2}{2 G_{N}}\chi(\Sigma_{\rt})+\frac{L^2}{8 G_{N}} \int\limits_{\Sigma_\rt} \diff^2 x \sqrt{\gamma}\mathcal{F},
\end{equation}
where L is the AdS radius. We show that the information on the shape deformation is controlled by the trace of the AdS curvature $\mathcal{F}$ of the RT surface. The first term is a topological contribution, given by the Euler characteristic $\chi(\Sigma_\rt)$, being non-local in the same way as free energy of the corresponding theory on $\mathbb{S}^3$.

In section \ref{sec:Willmore}, we derive the relation between renormalized EE and Willmore energy $\mathcal{W}$ of the doubled minimal surface $\Sigma_\rt$, given by
\begin{equation}
S^{\ren}_{\EE}\left (A\right ) = -\frac{L^{2}}{8G_N}\mathcal{W}\left (2\Sigma _{\rt}\right ),
\end{equation}
when the RT surface is embedded in AdS$_4$ bulk. This relation allows to map the strong subadditivity property of EE to a constraint on the AdS curvature. We show that the validity of the renormalized area formula holds for non-minimal surfaces, as well. 


\section{Renormalization of entanglement entropy from extrinsic counterterms}\label{Ren}


In this section, we review the cancellation of divergences that arise in the Einstein-Hilbert action when evaluated in asymptotically AdS (AAdS) spacetimes. We apply the extrinsic counterterms scheme, worked out in refs.\citep{Olea:2005gb,Olea:2006vd,Miskovic:2014zja,Miskovic:2009bm}. This produces a finite Euclidean action in order to obtain a renormalized entanglement entropy $S^{\text{ren}}_{\EE}$ by means of the relation \eqref{REdef}. In that respect, Kounterterms is a prescription alternative to standard holographic renormalization developed in refs.\citep{Emparan:1999pm,Kraus:1999di,deHaro:2000vlm,Balasubramanian:1999re,Henningson:1998gx,Papadimitriou:2004ap,Papadimitriou:2005ii}. 

Renormalized holographic EE has been computed for CFTs dual to Einstein-Hilbert gravity in an arbitrary dimension \citep{Anastasiou:2017xjr,Anastasiou:2018mfk,Anastasiou:2018rla,Anastasiou:2019ldc}. In these works, the universal contribution to EE is successfully extracted, removing all scheme-dependent quantities. In odd-dimensional CFTs, for spherical entangling surface, the renormalized EE corresponds to the free energy of a CFT residing on $\mathbb{S}^d$. In the case of even-dimensional CFTs, the only nonvanishing term is the logarithmic divergence, whose coefficient is the Weyl anomaly of the theory.


In what follows, we will restrict ourselves to odd $d$-dimensional CFTs, which correspond to even-dimensional dual gravity theories on an AAdS $(d +1)$-dimensional spacetime. The metric of this class of spacetimes is written in the Fefferman-Graham (FG) gauge as
\begin{equation} \label{Alads}
\diff s^{2} =G_{\mu \nu }dx^{\mu }\diff x^{\nu } =\frac{1}{z^{2}}\left (L^{2}\diff z^{2} +g_{ab}\left (z ,x\right )\diff x^{a}\diff x^{b}\right ) ,
\end{equation}
where $z$ is the holographic radial coordinate. The singularity at $z=0$, where the conformal boundary is located, induces a conformal structure at asymptotic infinity. The conformal boundary is endowed with a metric $g_{ab}\left (z ,x\right )$ which accepts an expansion of the form
\begin{equation}
g_{ab}\left (z ,x\right ) =g_{ab}^{\left (0\right )}\left (x\right ) +z^{2}g_{ab}^{\left (2\right )}\left (x\right ) +\ldots +z^{d}g_{ab}^{\left (d\right )}\left (x\right )  +z^{d}h_{ab}^{\left (d\right )}\left (x\right )\log \left (z^{2}\right ) +\ldots  .
\end{equation}

In the Kounterterms method, for even-dimensional manifolds $\mathcal{M}_{2n}$ with $2n=d+1$, the renormalized Einstein-AdS action $I^\text{ren}_\text{E}$ is achieved through the addition of the corresponding $n$-th Chern form $B_{2n-1}$, as
\begin{equation}\label{renB}
I^{\text{ren}}_{\text{E}}\left[\mathcal{M}_{2n}\right]=\frac{1}{16\pi G_{N}}\int\limits_{\mathcal{M}_{2n}}\diff^{2n}x\sqrt{|G|}(R-2\Lambda)+\frac{c_{2n}}{16\pi G_{N}}\int\limits_{\partial\mathcal{M}_{2n}}B_{2n-1},
\end{equation}
where the coefficient $c_{2n}$ is defined as
\begin{equation}\label{eq:cdplus1}
c_{2n}=(-1)^n\frac{L^{2n-2}}{n\Gamma(2n-1)},
\end{equation}
and the $n$-th Chern form reads
\begin{IEEEeqnarray}{ll}
\nonumber
B_{2n-1}&=-2n\int_0^1\diff t\sqrt{h}\updelta_{a_1\ldots a_{2n-1}}^{b_1\ldots b_{2n-1}}K_{b_1}^{a_1}\left(\frac{1}{2}\mathcal{\hat{R}}_{b_2b_3}^{a_2a_3}-t^2K_{b_2}^{a_2}K_{b_3}^{a_3}\right)\times\dots\\
&\dots\times\left(\frac{1}{2}\mathcal{\hat{R}}_{b_{2n-2}b_{2n-1}}^{a_{2n-2}a_{2n-1}}-t^2K_{b_{2n-2}}^{a_{2n-2}}K_{b_{2n-1}}^{a_{2n-1}}\right).
\end{IEEEeqnarray}
Here, $h_{ab}=g_{ab} \left(z,x\right)/z^2$ is the induced metric at a constant radius, $\mathcal{\hat{R}}_{cd}^{ab}$ is the intrinsic Riemann curvature tensor, $K^{a}_{b}$ the extrinsic curvature and $\updelta_{b_1\ldots b_{2n-1}}^{a_1\ldots a_{2n-1}}$ is the generalized Kronecker delta.

Note that the Euler theorem for manifolds with a boundary takes the form,
\begin{equation}
\int\limits_{\mathcal{M}_{2n}}\diff^{2n}x\sqrt{|G|}\mathcal{E}_{2n}=(4\pi)^{n}\Gamma\left(n+1\right)\chi\left[\mathcal{M}_{2n}\right]+\int\limits_{\partial \mathcal{M}_{2n}}B_{2n-1},
\end{equation}
expressing the equivalence of $B_{2n-1}$ with the topological term
\begin{equation}
\mathcal{E}_{2n}=\frac{1}{2^{n}}\updelta_{\mu_1\ldots\mu_{2n}}^{\nu_1\ldots\nu_{2n}}R^{\mu_1\mu_2}_{\nu_1\nu_2}\cdots R^{\mu_{2n-1}\mu_{2n}}_{\nu_{2n-1}\nu_{2n}},
\end{equation}
up to the Euler characteristic of the manifold $\chi\left[\mathcal{M}_{2n}\right]$. Using this result, we can rewrite expression \eqref{renB} exclusively in terms of bulk quantities as
\begin{equation}\label{IrenE}
I^{\text{ren}}_{\text{E}}=\frac{1}{16\pi G_{N}}\int\limits_{\mathcal{M}_{2n}}\diff^{2n}x\sqrt{|G|}(R-2\Lambda+c_{2n}\mathcal{E}_{2n})-\frac{(-1)^n}{4G_{N}}\frac{\pi^{(2n-1)/2}L^{2n-2}}{\Gamma[(2n-1)/2]}\chi\left[\mathcal{M}_{2n}\right].
\end{equation}
In ref.\citep{Anastasiou:2018rla}, it was shown that the quantity inside the integral in the above formula can be rewritten in terms of a polynomial of the tensor
\begin{equation} \label{FAds}
F^{\mu_1\mu_2}_{\nu_1\nu_2}=R^{\mu_1\mu_2}_{\nu_1\nu_2}+\frac{1}{L^2}\updelta^{\mu_1\mu_2}_{\nu_1\nu_2},
\end{equation}
known as \textit{AdS curvature}. In doing so, the action adopts the form
\begin{equation}\label{Iren}
I^{\text{ren}}_{\text{E}}=\frac{1}{16\pi G_{N}}\int\limits_{\mathcal{M}_{2n}}\diff^{2n}x\sqrt{|G|}L^{2n-2}P_{2n}(F)-\frac{(-1)^n}{4G_{N}}\frac{\pi^{(2n-1)/2}L^{2n-2}}{\Gamma[(2n-1)/2]}\chi\left[\mathcal{M}_{2n}\right].
\end{equation}
where the polynomial of the AdS curvature introduced reads
\begin{equation}\label{eq:Pol}
P_{2n}(F)=\frac{1}{2^{n}n\Gamma(2n-1)}\sum_{k=1}^n\frac{(-1)^k[2(n-k)]!2^{(n-k)}}{L^{2(n-k)}}\binom{n}{k}\updelta^{\nu_1\ldots\nu_{2k}}_{\mu_1\ldots\mu_{2k}}F^{\mu_1\mu_2}_{\nu_1\nu_2}\cdots F^{\mu_{2k-1}\mu_{2k}}_{\nu_{2k-1}\nu_{2k}}.
\end{equation}
The AdS curvature, of particular convenience in AdS gravity, measures the deviation of the space with respect to global AdS. Notice that the renormalized action consists on the addition of two terms: a topological one, given by the Euler characteristic of the manifold, and another one, characterized by the AdS curvature. This decomposition has been earlier found on the mathematical literature \citep{alexakis2010renormalized} in connection to the concept of renormalized volume. It reflects the equivalence between this quantity and renormalized EE, up to a proportionality constant that depends on the dimension of the manifold \citep{Anastasiou:2018rla}. This result will be of central importance afterwards for the renormalized EE of deformed entangling surfaces, as information on the deformation is entirely contained in the polynomial $P_{2n-2}(\mathcal{F})$.

Once we have the renormalized form of the Euclidean action $I^{\text{ren}}_{\text{E}}$, we evaluate it on the conically singular manifold $\mathcal{M}_{2n}^{(\alpha)}$ in order to use eq.\eqref{REdef}. Properties of curvature invariants defined on squashed cone manifolds like $\mathcal{M}_{2n}^{(\alpha)}$ have been developed in refs.\citep{Fursaev:1995ef,Mann:1996bi,Dahia:1998md,atiyah_lebrun_2013, Fursaev:2013fta}. For our purposes, we recall the relations
\begin{IEEEeqnarray}{ll}
R^{(\alpha)}&=R+4\pi(1-\alpha)\delta_\Sigma, \\
{R^{(\alpha)}}^{\mu\nu}_{\rho\sigma} &= R^{\mu\nu}_{\rho\sigma}+ 2\pi \left(1-\alpha\right) N^{\mu \nu}_{\rho\sigma} \delta_{\Sigma} 
\end{IEEEeqnarray}
where $N^{\mu \nu}_{\rho\sigma}=n^{\left(i\right)\mu}n\indices{^{\left(i\right)}_{\rho}} n\indices{^{\left(j\right)}^{\nu}}n\indices{^{\left(j\right)}_{\sigma}}-n^{\left(i\right)\mu}n\indices{^{\left(i\right)}_{\sigma}} n\indices{^{\left(j\right)}^{\nu}}n\indices{^{\left(j\right)}_{\rho}}$ is a linear combination of the $i$-th normal vector to the surface $\Sigma$, $n^{\left(i\right)\mu}$. Here $R^{(\alpha)}$ and ${R^{(\alpha)}}^{\mu\nu}_{\rho\sigma}$ denote the Ricci scalar and Riemann tensor evaluated at the orbifold, respectively. The unindexed tensors indicate the regular part of the corresponding bulk tensor and $\delta_\Sigma$ is a $(2n-2)$-dimensional Dirac delta localized at the conical singularity. As a consequence,
\begin{equation}
\int\limits_{\mathcal{M}_{2n}^{(\alpha)}}\diff^{2n}x\sqrt{G}\delta_\Sigma=\int\limits_{\Sigma}\diff^{2n-2}y\sqrt{\gamma},
\end{equation}
where $\Sigma$ is the codimension-2 locus of the conical singularity and $\gamma$ the induced metric on the $\Sigma$ hypersurface. We assigned the coordinate $y^{a}$ to parametrize the worldvolume of $\Sigma$. In ref.\citep{Anastasiou:2018rla}, it was shown that the Einstein-AdS action evaluated on the orbifold consists on the sum of a regular part and a term localized at the conical defect. The explicit form is
\begin{IEEEeqnarray}{ll}
\nonumber\label{eq:Iren}
I^{\text{ren}}_{\text{E}}\left[\mathcal{M}_{2n}^{(\alpha)}\right]&=\frac{L^{2n-2}}{16\pi G_{N}}\int\limits_{\mathcal{M}_{2n}^{(\alpha)}\setminus\Sigma}\diff^{2n}x\sqrt{|G|}P_{2n}(F)-\frac{(-1)^n}{4G_{N}}\frac{\pi^{(2n-1)/2}L^{2n-2}}{\Gamma[(2n-1)/2]}\chi\left[\mathcal{M}_{2n}^{(\alpha)}\setminus\Sigma\right]\\
&+\frac{1-\alpha}{4 G_{N}}\text{Vol}^{\text{ren}}\left[\Sigma\right]+\mathcal{O}\left[\left(1-\alpha\right)^2\right],
\end{IEEEeqnarray}
where $ \mathcal{M}_{2n}^{(\alpha)}\setminus\Sigma$ is identified as the regular manifold $\mathcal{M}_{2n}$ given by the $\alpha \rightarrow 1$ limit, and 
\begin{equation}\label{eq:Vol}
\text{Vol}^{\text{ren}}\left[\Sigma\right]=-\frac{L^{2n-2}}{2(2n-3)}\int\limits_{\Sigma}\diff^{2n-2}y\sqrt{\gamma}P_{2n-2}(\mathcal{F})-\frac{(-1)^n}{4G_{N}}\frac{\pi^{(2n-1)/2}L^{2n-2}}{\Gamma[(2n-1)/2]}\chi[\Sigma],
\end{equation}
is the renormalized volume of the codimension-2 manifold.
The factor proportional to the angular deficit
\begin{equation} \label{Tension}
T=\frac{1-\alpha}{4G},
\end{equation}
can be regarded as the cosmic brane tension of the Nambu-Goto action, in the interpretation given by Dong \citep{Dong:2016fnf}.

It is important to stress that the expression given for the renormalized volume in eq.\eqref{eq:Vol} is generic and not restricted to minimal hypersurfaces. In particular, for CFTs which are dual to Einstein-AdS gravity, when $\Sigma$ is minimal, corresponds to the RT surface.
When the limit $\alpha\rightarrow 1$ is taken in the renormalized action \eqref{eq:Iren}, the only surviving term \eqref{REdef} is the one coming from the Nambu-Goto action
\begin{equation}\label{eq:RTRen}
S^{\text{ren}}_{\EE} \left (A\right )=\frac{\text{Vol}^{\text{ren}}(\Sigma_{\rt})}{4 G_{N}},
\end{equation}
On the LHS, $A$ is a spatial entangling region in CFT$_{d}$ while on the RHS appearing the renormalized volume of the homologous RT surface $\Sigma _{\rt}$. 
Therefore, the computation of the renormalized entanglement entropy depends on AdS curvature and the Euler characteristic of the codimension-2 surface, attending to expression \eqref{eq:Vol}.

This calculation can be equivalently be interpreted as the renormalized volume of a tensionless codimension-2 brane $\Sigma $ embedded in a $2n$-dimensional AAdS Einstein spacetime, for a  minimal surface $\Sigma$ \citep{Dong:2016fnf}.

For a spherical entangling surface, the polynomial $P_{2n-2}\left(\mathcal{F}\right)$ vanishes identically. The contribution to the holographic EE is coming uniquely from the topology of the RT surface, which is an hemisphere. Because the Euler characteristic is $\chi[\Sigma_{\rt}]=1$, the finite part of the EE of a ball-shaped surface takes the form
\begin{equation}\label{eq:RenEgen}
S^{\text{ren}}_{\EE}=\frac{(-1)^{(d-1)/2}}{4 G_{N}}\frac{\pi^{d/2}L^{d-1}}{\Gamma(d/2)},
\end{equation}
where we have re-expressed the result in terms of the odd-dimensional $d$ of the CFT. Notice that this result is in agreement with the universal part of the EE \cite{Nishioka:2018khk}. As shown by Casini, Huerta and Myers in ref.\citep{Casini:2011kv},  $S^{\text{ren}}_{\EE}$ is equivalent to the free energy of a CFT$_d$ on a spherical background $\mathbb{S}^d$. This relation is of relevance for RG flows, as $F_{\mathbb{S}^d}$ is a monotonic function in $d=3$. Once the general picture has been discussed, we will illustrate explicitly the duality AdS$_4$/CFT$_3$ in this context by particular examples.

\subsection{Entanglement entropy in AdS$_4$/CFT$_3$ in the global coordinate patch}\label{sec:example}

The use of the extrinsic counterterms in the renormalization of holographic EE has been applied for spatial entangling regions embedded on a flat background, in the Poincar\'e-AdS patch, in the context of gauge/gravity duality \citep{Anastasiou:2017xjr}. In particular, in what follows, we study the EE of a polar cap-like entangling region immersed on an Einstein Static Universe background (ESU), \textit{i.e.}, $\mathbb{R} \times \mathbb{S}^2$ to account for properties of a CFT$_3$. In this case, the dual bulk geometry is given by global AdS$_4$ spacetime, whose line element reads

\begin{equation}
\diff s^2=-\left(1+\frac{r^2}{L^2}\right)\diff t^2+\left(1+\frac{r^2}{L^2}\right)^{-1}\diff r^2+r^2\diff\Omega_2^2,
\end{equation}
where $\diff\Omega_2^2=\diff \theta^2+\sin^2\theta\diff\phi^2$ is the metric of $\mathbb{S}^2$.

In the RT picture, the minimality condition for $\Sigma$, in order to be homologous to the circular entangling surface at the boundary, amounts to the vanishing of the trace of the extrinsic curvature $K_{ab}^{\left (i\right )}$ along whichever normal direction to $\Sigma$. Here, the label index $ i =\{1,2\}$ represents these directions. Indeed the equations of motion of the surface can be derived from the Nambu-Goto action. That, for the case of the Einstein-AdS gravity, results in the condition \citep{Bhattacharyya:2013sia,Bhattacharyya:2014yga}
\begin{equation}\label{eq:K0}
K^{(i)}=0.
\end{equation}
Considering that the two-dimensional orthogonal space is spanned along $i=t,r$, the induced metric $\gamma$ of $\Sigma$, is given by
\begin{equation}
\diff s^2_\gamma=-\left(1+\frac{r^2}{L^2}\right)\diff t^2+\left(r^2+\frac{L^2 {r'}^2}{L^2+r^2}\right)\diff \theta^2+r^2\sin^2\theta\diff\phi^2
\end{equation}
where we have parametrized the geometry with the embedding function $r=r(\theta)$ and $r'=\partial_\theta r(\theta)$.

Solving the second order differential equation that results from eq.\eqref{eq:K0}, we find that the RT surface is characterized by \citep{Hubeny:2007xt,Hubeny:2012wa,Bakas:2015opa} the function
\begin{equation}
r^2(\theta)=\frac{L^2\cos^2\theta_0}{\cos^2\theta\sin^2\theta_0-\sin^2\theta\cos^2\theta_0}.
\end{equation}
For this embedding, the polynomial $P_{2}\left(\mathcal{F}\right)$ in eq.\eqref{eq:Vol} vanishes identically, as it is a constant-curvature subspace. The only nonvanishing part is the topological one, for which the universal part of the EE takes the form
\begin{equation} \label{eq:Srenglobalads}
S^{\text{ren}}_{\EE}=-\frac{\pi L^2}{2 G_{N}}.
\end{equation}
Thus, even though this time the spherical entangling surface is  immersed in the curved background of ESU metric, eq.\eqref{eq:Srenglobalads} matches the one obtained for the flat case \citep{Anastasiou:2017xjr}.

\section{Renormalized entanglement entropy of a deformed disk}\label{RenEx}
In this section, we calculate the finite contribution to the EE of a spatial entangling region for a CFT$_{3}$ on the ground state. To this end, we consider a deformed disk whose dual geometry corresponds to global AdS$_4$. The universal part of the holographic EE for such region was first obtained in refs.  \citep{Allais:2014ata,Mezei:2014zla} for a  general class of gravity theories. Such result was later confirmed from field theory computations in ref.\cite{Faulkner:2015csl}.

We shall study the deformation in two coordinate systems:  polar coordinates (following \citep{Allais:2014ata}) and spherical coordinates (in order to make contact with refs.\citep{Mezei:2014zla} and \citep{Faulkner:2015csl}). Using the Kounterterms, we make contact with the renormalized volume of the RT surface \eqref{eq:Vol}, which contains both local (curvature) and global (topological) terms \citep{Anastasiou:2018mfk}. Our analysis below allows us to track the origin of the shape-dependent contributions to the curvature part in eq.\eqref{eq:Vol}.

\subsection{Deformed disk in polar coordinates} 
 
\begin{figure}
\centering
\includegraphics[width=0.8\textwidth]{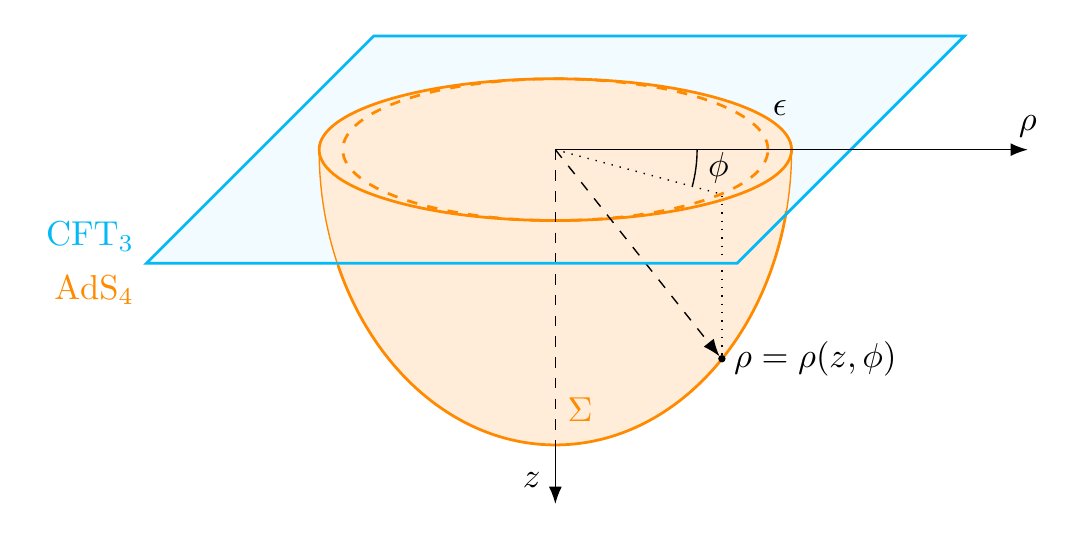}
\caption{Time slice of minimal co-dimension two surface $\Sigma$ with an elliptical deformation $\epsilon$ ($\ell=2$).}\label{fig1}
\end{figure}

Consider the Poincar\'e-AdS$_4$ spacetime, written in polar coordinates as
\begin{equation}\label{AdS4polar}
\diff s^2=\frac{L^2}{z^2}\left(-\diff t^2+\diff z^2+\diff \rho^2+\rho^2\diff\phi^2\right).
\end{equation} 
We define the embedding function of the RT surface $\Sigma_{\rt}$ by $\rho(z,\phi)$, where $\rho$ and $\phi$ are the radial and the angular coordinate at the boundary, respectively. The deformation breaks the azimuthal symmetry of $\Sigma_{\rt}$. Hence, the simplification used in the section \ref{sec:example} is not applicable. In this case, the codimension-2 induced metric reads\begin{equation}\label{indEAdS}
\diff s^2_\gamma=\frac{L^2}{z^2}\left[\left(1+{\rho'}^2\right)\diff z^2+\left(\rho^2+{{\dot{\rho}}}^2\right)\diff\phi^2+2\rho'\dot{\rho}\diff z\diff \phi\right],
\end{equation}
where $\rho'=\partial_z \rho(z,\phi)$ and $\dot{\rho}=\partial_\phi \rho(z,\phi)$. It is indeed easy to find the equations of motion of the RT surface following eq.\eqref{eq:K0}. If we consider the binormal directions as $i=t,r$, we find that
\begin{equation}
K\indices{^{(r)}_z^z}+K\indices{^{(r)}_\theta^\theta}=0,
\end{equation}
provided that the temporal foliation is constant, what implies into $K\indices{^{(t)}_z^z}=K\indices{^{(t)}_\theta^\theta}=0$. This leads to the equations of motion
\begin{equation}\label{eom}
\frac{\rho\left(1+\rho'^2\right)}{m z^2}-\partial_z\left(\frac{\rho^2\rho'}{m z^2}\right)-\frac{1}{z^2}\partial_\phi\left(\frac{\dot{\rho}}{m}\right)=0,
\end{equation}
where we have introduced an auxiliary function
\begin{equation}
m=m(z,\phi)=\sqrt{\rho^2\left(1+{\rho'}^2\right)+\dot{\rho}^2}\ .
\end{equation}
In absence of deformations, the embedding function \eqref{eom} is parametrized by a hemisphere of unit radius, $\rho^2=1-z^2$. The shape can be deformed as linear perturbations around the unitary circle of the form $\rho(\phi)=\left[1+\epsilon f(\phi)\right]$, where $\epsilon$ is the deformation parameter \citep{Allais:2014ata}. Altogether, we assume that its embedding in AdS$_4$ geometry is given by the ansatz
\begin{equation}\label{ansatz1}
\rho(z,\phi)=\sqrt{1-z^2}\left[1+\epsilon f(z,\phi)\right],
\end{equation}
for the separation of variables $f(z,\phi)=R(z)\Phi(\phi)$. The corresponding functions satisfy the conditions  $R(0)=1$ and $\Phi(\phi)=\Phi(\phi+2 \pi)$ at the conformal boundary. This is a consequence of the homologous constraint on the RT surface, as it is anchored to the conformal boundary $z=0$. An additional condition comes from the fact that the maximum reach of the embedding does not change when the RT surface is deformed, what leads to $R(1)=0$ \citep{Allais:2014ata,Hubeny:2012ry}. (see Figure \ref{fig1})

Solving eq.\eqref{eom} for $R(z)$ and $\Phi(\phi)$, we obtain
\begin{equation}\label{solution1}
\rho(z,\phi)=\sqrt{1-z^2}\left[1+\epsilon\sum_{\ell}\left(\frac{1-z}{1+z}\right)^{\ell/2}\frac{1+\ell z}{1-z^2} \left(a_\ell\cos(\ell\phi)+b_\ell\sin(\ell\phi)\right)\right],
\end{equation}
where $\ell$ is the degree of the harmonic function and labels the deformation with respect to the circle.

Once we have obtained the embedding function \eqref{solution1}, we are able to compute the renormalized EE for the perturbed circle by using eq.\eqref{eq:RTRen}. For Einstein gravity in four dimensions this reads
\begin{equation}\label{Sren2}
S_{\EE}^\ren \left(\mathbb{S}_\epsilon^1 \right)=-\frac{\pi L^2}{2 G_{N}}\chi(\Sigma_{\rt})+\frac{L^2}{8 G_{N}} \int\limits_{\Sigma_\rt} \diff^2 x \sqrt{\gamma}\mathcal{F},
\end{equation}
where $\mathcal{F}$ is the trace of the AdS curvature tensor defined in \eqref{eq:Pol}.
Replacing the embedding function \eqref{solution1} into eq.\eqref{Sren2}, we obtain
\begin{equation}\label{SEHe}
S_{\EE}^\ren \left(\mathbb{S}_\epsilon^1 \right)=-\frac{\pi L^2}{2 G_{N}}\left[1+\epsilon^2\sum_{\ell}\frac{\ell\left(\ell^2-1\right)}{4}(a^{2}_\ell+b^2_{\ell})+\mathcal{O}(\epsilon^4)\right],
\end{equation}
what is in agreement with the holographic computation for an arbitrary perturbation of a circle performed in ref.\cite{Allais:2014ata}.

\subsection{Deformed disk in spherical coordinates} 
\begin{figure}
\centering
\includegraphics[width=0.8\textwidth]{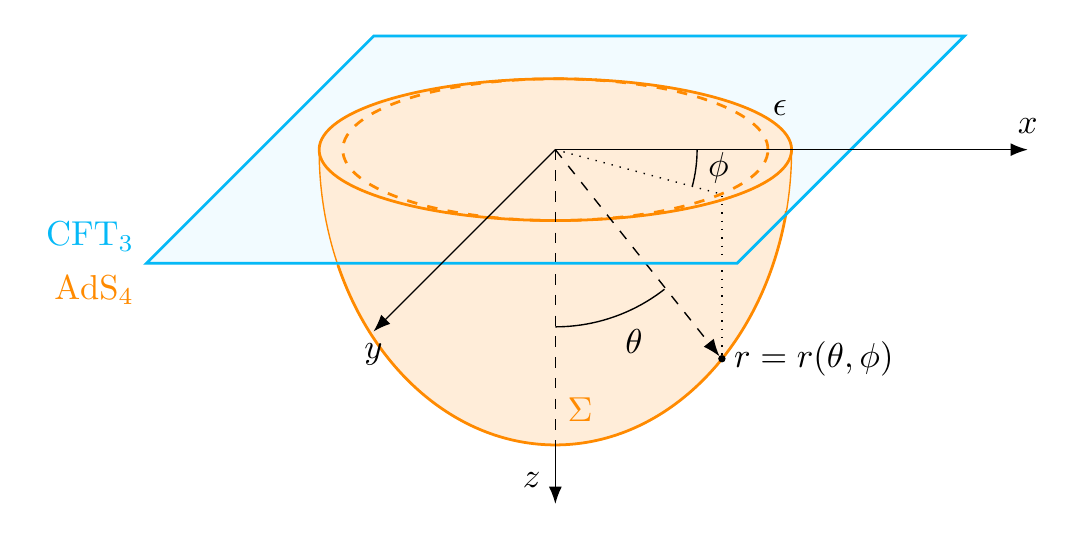}
\caption{Time slice of the  extremal co-dimension two surface $\Sigma$ with an elliptical deformation $\epsilon$ ($\ell=2$).}\label{fig2}
\end{figure}

Consider now the Poincar\'e-AdS spacetime written in spherical coordinates as
\begin{equation}
\diff s^2=\frac{L^2}{r^2\cos^2\theta}\left(-\diff t^2+\diff r^2+r^2\diff\theta^2+r^2\sin^2\theta\diff \phi^2\right).
\end{equation}
Polar and spherical coordinates are mapped into each other by the transformation
\begin{align}\label{sphericalcoords}
r=\sqrt{\rho^2+z^2},\quad
\theta=\arctan \frac{\rho}{z}.
\end{align}
In this coordinate system, the embedding function of the minimal surface $\Sigma_{\rt}$ is defined by $r=r(\theta,\phi)$, such that the induced metric is
\begin{equation}\label{gammaSphe}
\diff s^2_{\gamma}=\frac{L^2}{r^2\cos^2\theta}\left[\left(1+{r'}^2\right)\diff \theta^2+\left(1+{\dot{r}}^2\right)\diff\phi^2+2r'\dot{r}\diff\theta\diff\phi\right],
\end{equation}
where we denoted ${r'}=\partial_\theta r (\theta,\phi)$ and ${\dot{r}}=\partial_\phi r (\theta,\phi)$. The minimality condition \eqref{eq:K0} leads to the equation  for $r(\theta,\phi)$,
\begin{equation}\label{EOMsphe}
\frac{1}{m r^3\cos^2\theta}\left({r'}^2\sin^2\theta +\dot{r}^2\right)+\partial_\theta\left(\frac{r'\tan^2\theta }{r^2m}\right)+\frac{1}{\cos^2\theta}\partial_\phi\left(\frac{\dot{r}}{r^2 m}\right)=0,
\end{equation}
with the corresponding function
\begin{equation}
m=m(\theta,\phi)=\sin\theta\sqrt{1+\frac{{r'}^2+\dot{r}^2}{r^2}}.
\end{equation}
From eq.\eqref{EOMsphe}, in the undeformed case, the parametrization of the embedding function of the RT surface is given by the unit hemisphere, $r^2=1$. In a similar fashion as in the previous parametrization, we consider the linear perturbation of the entangling region as \citep{Mezei:2014zla}
\begin{equation}\label{ansatz2}
r(\theta,\phi)=1+\epsilon f(\theta,\phi).
\end{equation}
For a choice $f(\theta,\phi)=\Theta(\theta)\Phi(\phi)$, the boundary conditions correspond to a periodic function $\Phi$ with period $2 \pi$ and $\Theta\rightarrow 1$ at the conformal boundary, \textit{i.e.}, $\Theta(\frac{\pi}{2})=1$. Here, the maximal reach of the RT surface implies $\Theta(0)=0$ (see Figure \ref{fig2}). Thus, eq.\eqref{EOMsphe} for the ansatz \eqref{ansatz2} leads to a solution of the form 
\begin{equation}\label{embedding2}
r(\theta,\phi)=1+\epsilon\sum_\ell\tan^\ell\frac{\theta}{2}(1+\ell\cos\theta)\left[a_\ell\cos(\ell\phi)+b_\ell\sin(\ell\phi)\right]\,.
\end{equation}
For this embedding function, the nonvanishing AdS curvature component reads
\begin{equation}
\mathcal{F}\indices{^\theta^\phi_\theta_\phi}=-\epsilon^2\sum_\ell\frac{\ell^2(\ell^2-1)}{\pi L^2}\left(a_{\ell}+b_\ell\right)\tan\left(\frac{\theta}{2}\right)^{2\ell}\cot^4\left(\theta\right)+\mathcal{O}(\epsilon^3).
\end{equation}
Introducing this expression into eq.\eqref{Sren2}, yields
\begin{equation}
S_{\EE}^\ren \left(\mathbb{S}_\epsilon^1 \right)=-\frac{\pi L^2}{2 G_{N}}\left[1+\epsilon^2\sum_{\ell}\frac{\ell\left(\ell^2-1\right)}{4}(a^{2}_\ell+b^2_{\ell})+\mathcal{O}(\epsilon^4)\right].
\end{equation}
what matches exactly the result \eqref{SEHe} of the previous subsection, and, in turn, agrees with the formula for $d=3$ in ref.\citep{Mezei:2014zla}.


\subsection{Interpretation of the results}
A quick analysis of the results above leads to the fact that the $\mathcal{O} \left(\epsilon^2\right)$ contribution is coming only from the curvature part in formula \eqref{Sren2}. Indeed, the information on the deformation of the entangling region is only contained in the polynomial $\mathcal{P}_{2}(\mathcal{F})$. As it shall be discussed below, this behavior can be explained once the equivalence between the renormalized EE and the renormalized volume of the RT surface \eqref{eq:Vol}, is taken into account.

Continuous perturbations on the hemisphere do not modify its topology, leaving intact the Euler characteristic in eq.\eqref{eq:Vol}. As a consequence, its shape dependence is encoded only on the local properties of the manifold, which are reflected in the polynomial in the curvature ($\mathcal{F}$ term).


The term of the renormalized EE that is quadratic in the perturbation carries information on \textit{entanglement susceptibility}, associated to the change of shape of the entangling region \citep{Nozaki:2013wia,Nozaki:2013vta,Bhattacharya:2014vja,Faulkner:2015csl,Witczak-Krempa:2018mqx}. This quantity contains universal information due to the coefficient $C_{T}$ of the two-point correlation function of the energy-momentum tensor in a ground state of the  CFT$_3$. Indeed, the subleading term of the formula \eqref{SEHe}, can equivalently be written as
\begin{equation}
S^{\ren,\left(2\right)}_{\EE}(\mathbb{S}^1_\epsilon)=\frac{\pi^4 C_T}{24}\sum_\ell \ell(\ell^2-1)\left(a_\ell^2+b_\ell^2\right).
\end{equation}
This expression\footnote{In ref.\citep{Mezei:2014zla}, the proportionality constant differs by a factor $\pi$. This corresponds to a different normalization for the spherical harmonics, leading to an overall factor $1/\sqrt{\pi}$ for each one of them.}
 makes manifest the analogy between the entanglement susceptibility and $C_{T}$.
A posteriori, one can say that the AdS curvature of a deformed entangling region is a geometrical probe of $C_{T}$.

The leading-order contribution is a shape-independent constant that corresponds to the universal part of the EE of a circular entangling surface. This term is a topological number which is identified as the free energy of a CFT$_3$ in a $\mathbb{S}^3$ background, using gauge/gravity duality \citep{Casini:2011kv}.  As mentioned in the Introduction, the latter quantity provides a realization of the $F$-theorem. The matching to a notion of EE in terms of the Euler characteristic provides firmer ground to a connection between the topology and the effective number of degrees of freedom of the field theory.

\section{Renormalized volume and Willmore energy}\label{sec:Willmore}

\subsection{Minimal and non-minimal surfaces}


The connection between quantum information theoretic measures and geometry can be extended beyond EE. Dong in ref.\citep{Dong:2016fnf} showed that a similar area formula is valid for the calculation of the modular entropy. In this case, the codimension-2 hypersurface $\Sigma _{T}$ is not minimal, but its location is determined by the minimization of the Nambu-Goto action of a cosmic brane with tension $T$ \eqref{Tension}.

The prescription used in the present work for the cancellation of divergences in the holographic EE of entangling surfaces is linked to the volume renormalization given in the mathematical literature \citep{alexakis2010renormalized}. As shown in ref.\citep{Anastasiou:2018mfk}, isolating the finite contribution of the modular entropy amounts to the renormalization of the volume of $\Sigma _{T}$
\begin{equation}\label{renmodular}
\tilde{S}_m^\ren =\frac{\vol^\ren\left (\Sigma _{T}\right )}{4G_N}.
\end{equation}
Interestingly enough, both quantities, EE and modular entropy, are described by the same geometrical object, the renormalized volume of a codimension-2 hypersurface $\Sigma $. In 4D Einstein-AdS gravity, the corresponding renormalized volume of $\Sigma$ is given by \eqref{eq:Vol} and reads
\begin{equation} \label{renvolume}
\vol^\ren\left (\Sigma \right ) = -2\pi L^{2}\chi \left (\Sigma \right ) +\frac{^{}L^{2}}{4}\int\limits _{\Sigma }\diff^{2}y\sqrt{\gamma }\updelta _{ab}^{cd}\left (\mathcal{R}_{cd}^{ab} +\frac{1}{L^{2}}\updelta _{cd}^{ab}\right ).
\end{equation} 
This expression matches the renormalized area expression given in ref.\citep{Fischetti:2016fbh}. Notice that this formula holds whether $\Sigma$ is minimal or not.
A physical example of minimal surface is a soap film that spans between two wires. As there is no pressure difference between the sides, the membrane has zero mean curvature. In turn, soap bubbles are non-minimal, due to the difference of pressure at the interface \cite{isenberg1978science, reilly1982mean}. In the latter case, they are constant mean curvature surfaces, and modelled by Helfrich energy \cite{hopf2003differential}.


For extremal surfaces, minimality condition amounts to the vanishing of the trace of the extrinsic curvature  of the surface $\Sigma$ \eqref{eq:K0}. For this reason, it is useful to rewrite eq.\eqref{renvolume} in terms of $K_{ab}^{\left (i\right )}$ using the Gauss-Codazzi relation, for codimension-2 surfaces
\begin{equation}\label{gausscodazzi}
R_{cd}^{ab} =\mathcal{R}_{cd}^{ab} -{K^{\left (i\right )}}_{c}^{a}{K^{\left (i\right )}}_{d}^{b} +{K^{\left (i\right )}}_{d}^{a}{K^{\left (i\right )}}_{c}^{b} .
\end{equation}
Taking into account the antisymmetry of Kronecker delta, we find that
\begin{equation}
\vol^\ren\left (\Sigma \right ) = -2\pi L^{2}\chi \left (\Sigma \right ) +\frac{^{}L^{2}}{4}\int \limits_{\Sigma }\diff^{2}y\sqrt{\gamma }\updelta _{ab}^{cd}\left [\left .R_{cd}^{ab} +2{K^{\left (i\right )}}_{c}^{a}{K^{\left (i\right )}}_{d}^{b} +\frac{1}{L ^{2}}\updelta _{cd}^{ab}\right .\right ] .
\end{equation}
In addition, the Weyl tensor for Einstein-AdS spaces can be written as
\begin{equation}
W_{\mu \nu }^{\alpha \beta } =R_{\mu \nu }^{\alpha \beta } +\frac{1}{L^{2}}\updelta _{\mu \nu}^{\alpha \beta} ,
\end{equation}
what allows us to express the renormalized volume of $\Sigma$ as
\begin{equation} \label{renvol1}
\vol^{\ren}\left (\Sigma \right ) = -2\pi L^{2}\chi \left (\Sigma \right ) +\frac{L^{2}}{4}\int\limits_{\Sigma }\diff^{2}y\sqrt{\gamma }    \updelta _{ab}^{cd}\left [W_{cd}^{ab} +2{K^{\left (i\right )}}_{c}^{a}{K^{\left (i\right )}}_{d}^{b}\right ] .
\end{equation}
In turn, the extrinsic curvature in eq.\eqref{renvol1} can be decomposed into its trace $K^{\left (i\right )}$ and a traceless part $P^{\left(i\right)}_{ab}$ as
\begin{equation}
K\indices{^{\left (i\right )}_a_b} =P^{\left(i\right)}_{ab} +\frac{1}{2}\gamma _{ab}K^{\left (i\right )}.
\end{equation}
We can also replace the trace by the mean curvature $H^{(i)}$, which expresses a linear combination of the eigenvalues of $K\indices{^{\left (i\right )}_a_b}$, that is, $H^{\left (i\right )} =K^{\left (i\right )}/2$. Armed with these tools, we deduce that the renormalized volume in eq.\eqref{renvolume} can be equivalently written as
\begin{equation}\label{renvolnonminimal}
\vol^{\ren}\left (\Sigma \right ) = -2\pi L^{2}\chi \left (\Sigma \right ) +\frac{L^{2}}{2}\int\limits_{\Sigma }\diff^{2}y\sqrt{\gamma }\left[W_{ab}^{ab} +2{H^{(i)}}^{2} -P^{\left(i\right) a}_{b}P^{\left(i\right) b}_{a} \right]. 
\end{equation}
When a minimal two-dimensional surface $\Sigma_{\text{min}}$ is considered, the above relation reduces to
\begin{equation}\label{renvolminimal}
\vol^{\ren}\left (\Sigma_{\text{min}} \right ) = -2\pi L^{2}\chi \left (\Sigma_{\text{min}} \right ) +\frac{^{}L^{2}}{2}\int\limits_{\Sigma_{\text{min}} }\diff^{2}y\sqrt{\gamma }\left [W_{ab}^{ab} -P^{\left(i\right) a}_{b}P^{\left(i\right) b}_{a} \right ].
\end{equation}
The last two equations are in agreement with the result of Alexakis and Mazzeo in ref.\citep{alexakis2010renormalized} for the renormalized area of submanifolds. These equivalent expressions allow us to tell between the two prescriptions in eqs.\eqref{eq:RTRen} and \eqref{renmodular}: the RT surface satisfies the minimality condition while the cosmic brane used in modular entropy not. Renormalized area relations in eqs.\eqref{renvolnonminimal} and \eqref{renvolminimal} can be further simplified when considering entangling regions for a vacuum CFT. Since its gravity dual is global AdS$_{4}$ spacetime, which is conformally flat, the bulk Weyl tensor vanishes identically.

\subsection{Renormalized area and Willmore energy}
The relation between renormalized EE and the renormalized area of the RT surface, has two key ingredients. On one hand, the topology of the minimal surface, expressed by the Euler characteristic, which captures global properties of $\Sigma _{\rt}$. On the other hand, the local properties of $\Sigma _{\rt}$ are dictated by the AdS curvature term inside the integral in eq.\eqref{FAds}.

According to the analysis in section \ref{RenEx}, the deformation in the shape of a disk entangling region is encoded only at the curvature part of the renormalized EE, leaving the topological contribution unchanged.

A functional with similar properties, called the Willmore energy, has been introduced in mathematical literature \citep{marques2014willmore,willmore1996riemannian,toda}. It is defined for a smooth, closed and orientable surface $X$ embedded in $\mathbb{}\mathbb{R}^{3}$  and adopts the form,
\begin{equation}\label{willmore functional}
\mathcal{W}\left (X\right ) =\int\limits_{X}H^{2}\diff S , 
\end{equation}
where $H$ is the mean curvature of $X$ and $\diff S$ is the area element of the 2D metric. Furthermore, it acquires a minimal value when evaluated on spherical surfaces
\begin{equation} \label{willmore bound}
\mathcal{W}\left (X\right ) \geq 4\pi  .
\end{equation}
Therefore, it measures the deviation of $X$ from sphericity. It was conjectured by Willmore that a new bound arises when the genus of $X$ changes from zero to one. According to this, the functional for the $g=1$ submanifold has a bound
\begin{equation} \label{willmoreboundtorus}
\mathcal{W}\left (X_{g=1}\right ) \geq 2\pi^2.
\end{equation}
The conjecture was proved recently in ref.\citep{fern2012minmax}. These properties are fundamental for the analysis below, where 
we establish the connection between Willmore energy and quantum information theoretic measures, through the concept of renormalized volume of the entangling surface.

Willmore energy is defined for a closed surface. Therefore, the first obstacle is the fact formula \eqref{renvolume} involves an open two-dimensional surface anchored to the boundary of an AAdS space.

\begin{figure}
\includegraphics[width=.45\linewidth,valign=c]{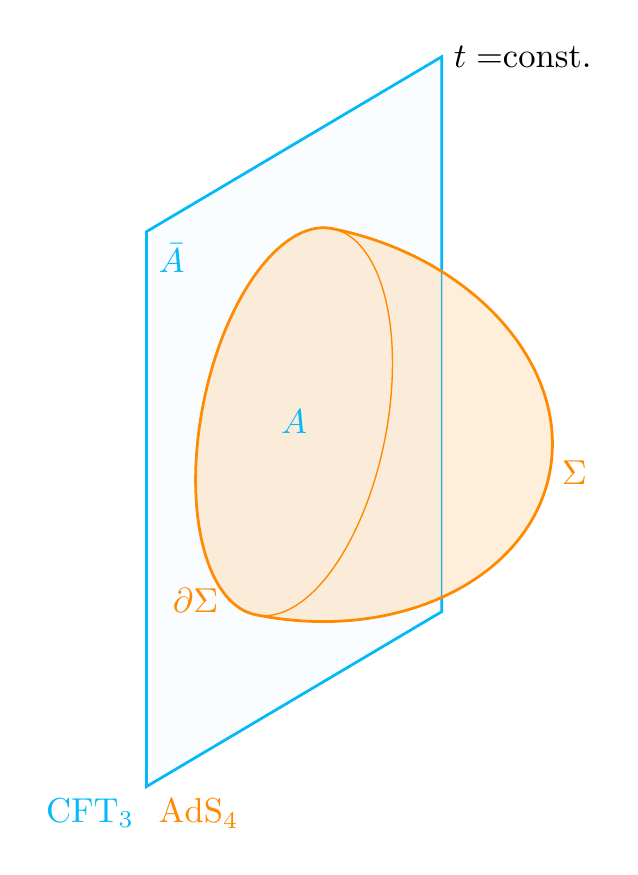} 
\hfill\vline\hfill
\begin{minipage}{.54\linewidth}
\centering
\includegraphics[width=.9\linewidth]{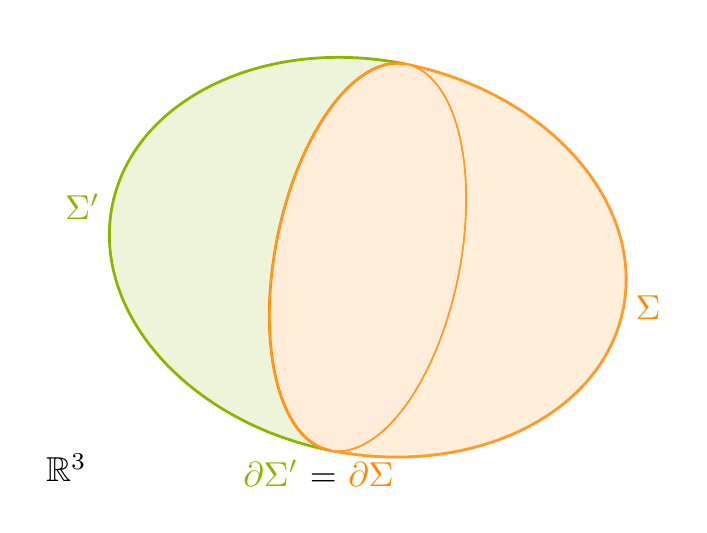}
\end{minipage}
\caption{The doubling of the minimal surface $\Sigma$ is achieved by gluing a copy $\Sigma'$ so that they are cobordant.}\label{fig:will}
\end{figure}

In order to overcome this problem, we generalize the field doubling method proposed in ref.\citep{Fonda:2015nma}, for AAdS manifolds. In this case, we consider the embedding of the codimension-2 surface $\Sigma $ and its reflection with respect to the $z =0$ plane, $\Sigma ^{ \prime }$. The intersection of $\Sigma $ and $\Sigma ^{ \prime }$ is the entangling curve $ \partial A$, at the conformal boundary, such that $\partial\Sigma=\partial\Sigma'=\partial A$. Continuity conditions at the interface situated at $z =0$, require the two surfaces to be immersed in a regular spacetime. In this case, its union  produces a closed two-dimensional surface $2 \Sigma=\Sigma \cup \Sigma'$, which is embedded into the smooth spacetime $\tilde{G}_{\mu \nu }$. A pictorial representation of the method is shown in Figure \ref{fig:will}.

With this geometrical setup in mind, we examine the rescaling properties of the renormalized volume of $\Sigma $ \eqref{renvolnonminimal} under generic Weyl transformations of the ambient metric $G_{\mu \nu } =\e{2\varphi }\mathcal{}\tilde{G}_{\mu \nu }$. Notice that, when $\varphi  = -\log(z/L)$, one recovers the eq.\eqref{Alads}. The Euler characteristic is a topological invariant and does not change under metric rescalings. Thus, we focus on the quantities which appear under the integral symbol in eq.\eqref{renvolnonminimal}.

By definition of the bulk line element in terms of the codimension-2 metric $\gamma_{ab}$
\begin{equation}
G^{\mu \nu } =\left (n\indices{^{\left (i\right )}^{\mu }}n\indices{^{\left (i\right )}^{\nu }} +e_{a}^{\mu }e_{b}^{\nu }\gamma ^{ab}\right ) ,
\end{equation}
where $n\indices{^{\left (i\right )}^{\mu }}$ are the corresponding normal vectors and $e_{a}^{\mu }$ are the frame vectors, we have that
\begin{equation}
n\indices{^{\left (i\right )}_{\mu }}=\e{\varphi }\tilde{n}\indices{^{\left (i\right )}_{\mu }}, \quad
\gamma _{ab} =\e{2\varphi }\tilde{\gamma }_{ab}.
\end{equation}
Here, the quantities with tilde indicate an embedding with respect to the regular metric $\tilde{G}_{\mu \nu }$.

On the other hand, the extrinsic curvature of $\Sigma $ transforms as
\begin{equation}\label{K}
K\indices{^{\left (i\right )}_a_b} =\e{\varphi }\left (\tilde{K}\indices{^{\left (i\right )}_a_b} +\tilde{\gamma }_{ab}\tilde{n}^{\left (i\right )} \cdot \partial \varphi\right ), 
\end{equation}
where we have omitted the indices in the contraction $\tilde{n}\indices{^{\left (i\right )}^{c}} \partial _{c}\varphi$. This expression allows us to write down the trace of $K\indices{^{\left (i\right )}_a_b}$ as
\begin{equation}\label{trK}
K^{\left (i\right )} =\gamma ^{ab}K\indices{^{\left (i\right )}_a_b} =\e{ -\varphi }\left (\tilde{K}^{\left (i\right )} +2\tilde{n}^{\left (i\right )}\partial \varphi \right ) , 
\end{equation}
and its traceless part as
\begin{equation} \label{tracelesK}
P^{\left(i\right)}_{a b} =\e{-\varphi } \tilde{P}^{\left(i\right)}_{a b} .
\end{equation}
A Weyl transformation of the area element is given by
$\diff^{2}y\sqrt{\gamma } =\diff^{2}y\sqrt{\tilde{\gamma }}e^{2\phi }$.
Then, it is straightforward to show that the following object is Weyl invariant
\begin{equation}\label{tracelessKweyl}
\int\limits_{\Sigma }\diff^{2}y\sqrt{\gamma }P^{\left(i\right) a}_{b} P^{\left(i\right) b}_{a} =\int\limits_{\Sigma }\diff ^{2}y\sqrt{\tilde{\gamma }}\tilde{P}^{\left(i\right) a}_{b} \tilde{P}^{\left(i\right) b}_{a} . 
\end{equation}
In turn, the square of the trace of the extrinsic curvature is not Weyl invariant
\begin{equation}\label{meancurvatureweyl}
\int\limits_{\Sigma }\diff^{2}y\sqrt{\gamma }{K^{\left (i\right )}}^{2} =\int \limits_{\Sigma }\diff^{2}y\sqrt{\tilde{\gamma }}\left [{{}\tilde{K}^{\left (i\right )}}^{2} +4\tilde{K}^{\left (i\right )} \left(\tilde{n}^{\left (i\right )} \cdot \partial\varphi \right) + 4\left (\tilde{n}^{\left (i\right )} \cdot \partial\varphi \right )^{2}\right ] .
\end{equation}
Altogether, the bulk Weyl tensor satisfies $W\indices{^a_b_c_d} =\tilde{W}\indices{^a_b_c_d}$, what implies the relation
\begin{equation}
W_{cd}^{ab} =\e{ -2\varphi }\tilde{G}^{bm}\tilde{W}\indices{^a_m_c_d} =\e{ -2\varphi }\tilde{W}_{cd}^{ab}.
\end{equation}
In doing so, the integral of the double subtrace of the Weyl tensor on the area element is proved to be invariant
\begin{equation}\label{weylsmooth}
\int\limits_{\Sigma }\diff^{2}y\sqrt{\gamma }W_{ab}^{ab} =\int\limits_{\Sigma }\diff^{2}y\sqrt{\tilde{\gamma }}\tilde{W}_{ab}^{ab} . 
\end{equation}
Therefore, the renormalized volume \eqref{renvolume}, when expressed in terms of the smooth metric, reads
\begin{IEEEeqnarray}{ll}
\nonumber\label{renarea1}
\vol^{\ren}\left (\Sigma \right ) =&\frac{^{}L^{2}}{2}\int\limits_{\Sigma }\diff^{2}y\sqrt{\tilde{\gamma }}\left[\tilde{W}_{ab}^{ab} +2{{}\tilde{H}^{\left (i\right )}}^{2} -\tilde{P}^{\left(i\right) a}_{b} \tilde{P}^{\left(i\right) b}_{a} +2\tilde{K}^{\left (i\right )}\left( \tilde{n}^{\left (i\right )} \cdot \partial\varphi \right)  +2\left(\tilde{n}^{\left (i\right )} \cdot \partial\varphi \right )^{2}\right]- \\
&-2\pi L^{2}\chi \left (\Sigma \right ) .
\end{IEEEeqnarray}
This formula adopts a more compact form by taking Gauss-Codazzi eq.\eqref{gausscodazzi} and the relations between bulk and codimension-2 curvature tensors \citep{Anastasiou:2019ldc}
\begin{IEEEeqnarray}{rl}
\tilde{P}^{\left(i\right) a}_{b} \tilde{P}^{\left(i\right) b}_{a}& ={{}\tilde{K}^{\left (i\right )}}_{b}^{a}{{}\tilde{K}^{\left (i\right )}}_{a}^{b}-2{{}\tilde{H}^{\left (i\right )}}^{2} , \notag \\
{{}\tilde{K}^{\left (i\right )}}_{b}^{a}{{}\tilde{K}^{\left (i\right )}}_{a}^{b}&=\tilde{R} +\tilde{R}_{abcd}n\indices{^{(i)}^a}n\indices{^{(i)}^c}n\indices{^{(j)}^b}n\indices{^{(j)}^d}  -2\tilde{R}_{ab}n\indices{^{(i)}^a}n\indices{^{(i)}^b} -\tilde{\mathcal{R}} +4{{}\tilde{H}^{\left (i\right )}}^{2} , \notag \\
\tilde{W}_{ab}^{ab}&=\tilde{R}_{ab}^{ab} -2\tilde{S}_{a}^{a} , \notag \\
\tilde{R}_{ab}^{ab}&=\tilde{R} +\tilde{R}_{abcd}n\indices{^{(i)}^a}n\indices{^{(i)}^c}n\indices{^{(j)}^b}n\indices{^{(j)}^d}  -2\tilde{R}_{ab}n\indices{^{(i)}^a}n\indices{^{(i)}^b} \notag,
\end{IEEEeqnarray}
where $\tilde{S}_{\nu }^{\mu }$ is the Schouten tensor of $\tilde{G}_{\mu \nu }$. Combining these expressions, we find that
\begin{equation}
\tilde{P}^{\left(i\right) a}_{b} \tilde{P}^{\left(i\right) b}_{a}=\tilde{W}_{ab}^{ab} +2\tilde{S}_{a}^{a} -\mathcal{R} +2{{}\tilde{H}^{\left (i\right )}}^{2},
\end{equation}
what leads to
\begin{equation}
\vol^{\ren}\left (\Sigma \right ) = -2\pi L^{2}\chi \left (\Sigma \right ) +L^{2}\int\limits_{\Sigma }\diff^{2}y\sqrt{\tilde{\gamma }}\left [\frac{1}{2}\mathcal{R} +\tilde{K}^{\left (i\right )} \left(\tilde{n}^{\left (i\right )} \cdot \partial\varphi  \right) +\left(\tilde{n}^{\left (i\right )} \cdot \partial\varphi \right )^{2}-\tilde{S}_{a}^{a}\right ] . \label{volsigma}
\end{equation}
Until now, we have treated the two-dimensional sheet $\Sigma $ as an open surface anchored to the $z =0$ plane. As discussed above, $\Sigma $ is also half of the closed surface $2\Sigma $, we have
\begin{equation}\label{vol2sigma}
\vol^{\ren}\left (\Sigma \right ) = -2\pi L^{2}\chi \left (\Sigma \right ) +\frac{^{}L^{2}}{2}\int\limits_{2\Sigma }\diff^{2}y\sqrt{\tilde{\gamma }}\left [\frac{1}{2}\mathcal{R} +\tilde{K}^{\left (i\right )}\left(\tilde{n}^{\left (i\right )} \cdot \partial\varphi \right) +\left(\tilde{n}^{\left (i\right )} \cdot \partial\varphi \right )^{2}-\tilde{S}_{a}^{a}\right ] . 
\end{equation}
For the compact manifold $2\Sigma$, the Euler theorem in two dimensions states that
\begin{equation}\label{eulertheorem}
\int\limits_{2\Sigma }\diff^{2}y \sqrt{\tilde{\gamma }} \mathcal{R} =4\pi \chi \left (2\Sigma \right ) , 
\end{equation}
as for the Euler characteristic, the relation
\begin{equation}\label{eulercharacteristic}
\chi \left (2\Sigma \right ) =2\chi \left (\Sigma \right ) ,
\end{equation}
holds without loss of generality.
With all the above equations, we have
\begin{equation}\label{renvolgeneric}
\vol^{\ren}\left (\Sigma \right ) =\frac{^{}L^{2}}{2}\int\limits_{2\Sigma }\diff^{2}y\sqrt{\tilde{\gamma }}\left [\tilde{K}^{\left (i\right )} \left(\tilde{n}^{\left (i\right )} \cdot \partial\varphi \right) +\left(\tilde{n}^{\left (i\right )} \cdot \partial\varphi \right )^{2} -\tilde{S}_{a}^{a}\right ] ,
\end{equation}
what is equivalent to the renormalized area formula in eq.\eqref{renvolume}. As a matter of fact, it is more general as it is valid for both minimal and non-minimal surfaces embedded in an AAdS$_{4}$ spacetime. Any constraint on the shape of $\Sigma $ can be readily implemented as a relation between the different terms in eq.\eqref{renvolgeneric}.

For instance, minimality condition leads to
\begin{equation}\label{minimality condition}
K^{\left (i\right )} =0 \Rightarrow \tilde{K}^{\left (i\right )} = -2\tilde{n}^{\left (i\right )} \cdot \partial\varphi  ,
\end{equation}
when written in terms of the smooth metric $\tilde{G}_{\mu \nu }$.
As a consequence, when a minimal surface $\Sigma _{\text{min}}$ is considered, the renormalized volume reduces to
\begin{equation}
\label{renvolsmoothminimal}
\vol^{\ren}\left (\Sigma _{\text{min}}\right ) = -\frac{^{}L^{2}}{2}\int\limits _{2\Sigma _{\text{min}}}d^{2}y\sqrt{\tilde{\gamma }}\left [{{}\tilde{H}^{\left (i\right )}}^{2} +\tilde{S}_{a}^{a}\right ] . 
\end{equation}
In particular, when $\Sigma _{\text{min}}$ is a spatial subregion of an AdS$_{4}$ spacetime, which corresponds to the conditions $\tilde{H}^{\left (t\right )} =0$ and $\tilde{S}_{a}^{a} =0$ ($\tilde{G}_{\mu \nu }$ is a locally flat space), the last equation reads
\begin{equation}\label{renvolwillmore}
\vol^{\ren}\left(\Sigma _{\text{min}}\right ) = -\frac{^{}L^{2}}{2}\int\limits_{2\Sigma _{\text{m}}}\diff^{2}y\sqrt{\tilde{\gamma }}\tilde{H}^{2} = -\frac{^{}L^{2}}{2}\mathcal{W}\left (2\Sigma _{\text{min}}\right). 
\end{equation}
Therefore, it is made explicit the connection between the renormalized area and the Willmore energy. The equivalence of these two geometric concepts lead to interesting inequalities about the AdS curvature of the minimal surface.

Indeed, when the closed surface $2\Sigma _{\text{min}}$ belongs to the topological class of the sphere ($g=0$), the combination of eqs.\eqref{renvolume},\eqref{renvolwillmore} and \eqref{eulercharacteristic} leads to the inequality
\begin{equation} \label{EEinequality}
\int\limits_{\Sigma _{\text{min}}}\diff^{2}y\sqrt{\gamma }\mathcal{F} \leq 0.
\end{equation}
For a toroidal closed surface ($g=1$), the inequality in eq.\eqref{willmoreboundtorus} gives
\begin{equation} \label{EEinequality2}
\int\limits_{\Sigma _{\text{min}}}\diff^{2}y\sqrt{\gamma }\mathcal{F} \leq -2\pi^2 .
\end{equation}
In this geometry, the bound is saturated by the Clifford torus \cite{Astaneh:2014uba}. Notice that in both cases the integral of the trace of the AdS curvature is non-positive.

From a different starting point, Alexakis and Mazzeo arrived at the same type of inequalities in ref.\citep{alexakis2010renormalized}.
Note that our derivation of the Willmore energy from renormalized volume relies on the existence of an AdS bulk. Thus, the bounds \eqref{EEinequality} 
and \eqref{EEinequality2} cannot be extended to other backgrounds. A generalization of these results to 
generic bulk and boundary geometries can be seen in ref.\citep{Fischetti:2016fbh}.

\section{Holographic entanglement entropy and Willmore energy}

The connection between renormalized area and Willmore energy provide us insight on surfaces immersed in a higher-dimensional manifold. In particular, the dependence of EE on the geometry becomes manifest when taking a minimal surface $ \Sigma_{\text{min}}$.
 
For RT surfaces, the renormalized area of $\Sigma_{\rt}$ is equivalent to the renormalized EE of the subregion $A$, and the following formula holds
\begin{equation} \label{EEwilmore}
S^{\ren}_{\EE}\left (A\right ) = -\frac{L^{2}}{8G_N}\mathcal{W}\left (2\Sigma _{\rt}\right ).
\end{equation}
One can map the universal part of EE of an entangling region for a vacuum state of the CFT$_3$ to the Willmore energy of a closed geometry constructed by gluing two copies of the RT surface.

Inequalities \eqref{willmore bound} and \eqref{willmoreboundtorus} set a bound to renormalized EE \eqref{EEwilmore}. For a doubled RT surface which correspond to $g=0$ \citep{Fonda:2015nma}, we get
\begin{equation}\label{renEEbound}
S^{\ren}_{\EE}\left (A\right ) \leq  -\frac{\pi L^{2}}{2G_N} ,
\end{equation}
that means the finite part of the EE is maximized for a circular surface among all the possible shapes within the same topological class. The same bound was obtained in ref.\citep{alexakis2010renormalized}.

In a similar fashion, when the closed surface $2\Sigma _{\rt}$ is of genus $g=1$, the finite term of the EE satisfies
\begin{equation}\label{renEEboundtorus}
S^{\ren}_{\EE}\left (A\right ) \leq  -\frac{\pi^2 L^{2}}{4G_N} .
\end{equation}
Therefore, the sphere results as the global maximum of the EE between surfaces of genus up to one. Astaneh, Gibbons and Solodukhin arrived at the same conclusion by extending their study to higher dimensional surfaces of genus larger than one in ref.\citep{Astaneh:2014uba}.

In section \ref{RenEx} we showed that a measure of the deformation of an entangling surface is given by the trace of the AdS curvature, subjected to the inequalities \eqref{EEinequality} and \eqref{EEinequality2}. This quantity is a holographic geometric probe of entanglement susceptibility in the dual CFT. Interestingly enough, the susceptibility is negative as a consequence of the strong subadditivity property of EE \citep{Nozaki:2013vta,Faulkner:2015csl,Witczak-Krempa:2018mqx}. Hence, strong subadditivity imposes a restriction on the curvature of the RT surface side which reads
\begin{equation} \label{geometricSSA}
\int\limits_{\Sigma _{\rt}}\diff^{2}y\sqrt{\gamma } \left(\mathcal{R} + \frac{2}{L^2} \right) \leq 0 .
\end{equation}
An analogous constraint on the spacetime curvature was derived in ref.\citep{Bhattacharya:2014vja} in the context of covariant EE in AdS$_3$/CFT$_2$.

\subsection{$F$-theorem and Willmore energy}

The universal term $s_\text{univ}$ of EE of a disk-like entangling region is a relevant quantity, as it is identified with the free energy $F_{\mathbb{S}^3}$ of a CFT$_{3}$ on a spherical background $\mathbb{S}^{3}$ \citep{Casini:2011kv}. In addition, $F_{\mathbb{S}^3}$ has been proven to be a $F$-function along the RG flows in $d=3$ \citep{Jafferis:2011zi,Myers:2010xs}, what reflects the degrees of freedom of the theory. 

Consider the EE of a spatial subregion $A$ for a 3-dim CFT
\begin{equation} \label{EEexpansion}
S_{\EE} \left (A\right ) =\frac{\text{Area}\left (\partial A\right )}{\delta} -s_{\text{univ}}\left (A\right ),
\end{equation}
where $\delta$ is the regulator in eq.\eqref{Seven}. By an adequate manipulation of eqs.\eqref{EEwilmore} and \eqref{EEexpansion}, it is straightforward to show that
\begin{equation}
s_\text{univ}\left (A\right) =\frac{L^{2}}{8G_N}\mathcal{W}\left (2\Sigma _{\rt}\right ) , \label{Fwillmore}
\end{equation}
whereas combining eqs.\eqref{renEEbound} and \eqref{EEexpansion}, one can arrive at the inequality
\begin{equation}
s_\text{univ}\left (A\right) \geq \frac{\pi L^{2}}{2G_N} = F_{\mathbb{S}^3}. \label{Finequality}
\end{equation}

One can assume that the Casini-Huerta-Myers (CHM) map linking the finite term in the EE and the number of degrees of freedom of the theory is valid for any shape of the entangling surface. On the other hand, the Wilsonian picture of the RG flows indicates that the microscopic degrees of freedom depend on the energy of the theory. In this picture, the energy acquire global characteristics as it affects the degrees of freedom independently on their position and local properties on the manifold. Therefore, $s_\text{univ}\left (A\right )$ should depend on global features of the entangling surface, namely the topological contribution of the renormalized EE in eq.\eqref{EEwilmore}.

Furthermore, for surfaces within the same topological class only the ones having the maximum renormalized area are adequate probes of the degrees of freedom of the theory. This can be seen as coming from their maximum capacity of information storage. Due to the fact that the circle is the global maximum of the area among all 2D geometries of different genus --- as shown in eqs.\eqref{renEEbound} and \eqref{renEEboundtorus} --- it represents a strong candidate to a proper measure of the degrees of freedom.

In addition, surfaces of maximal renormalized area with $g > 0$, are entangling regions that cannot fully cover a spatial slice of the CFT in their maximum extension. Physically, that means that they cannot account for all the degrees of freedom of the theory. In turn, entangling regions of $g=0$ can potentially cover the full manifold in its totality. Indeed, the fact that the disk has the maximum renormalized area in the $g=0$ topological class, implies that is able to encode all the information  in the theory \footnote{We thank I. J.  Araya for comments on this point}.

\section{Discussion}

In the present paper, we have studied the shape dependence of entanglement entropy in 3-dim CFTs which are holographically dual to Einstein-AdS gravity. The finite part of the entanglement entropy is expressed as the renormalized volume of the RT surface \eqref{eq:RTRen} for CFTs in odd dimensions. It consists on two contributions: a topological part, proportional to the Euler characteristic of $\Sigma$ (shape independent); and a curvature term, which encodes the information of the deformation of the entangling region with respect to a constant-curvature condition.

We have presented explicit computations on entangling regions with deformations for 3-dimensional CFTs, along the line of refs.\citep{Anastasiou:2017xjr,Anastasiou:2018mfk,Anastasiou:2018rla}. We match the results found in the literature given in refs.\citep{Mezei:2014zla,Allais:2014ata}. Our analysis shows that the number of degrees of freedom of the field theory is given by the topological part. In turn, the quadratic term in the deformation is coming from the integral of the AdS curvature. This means, that the AdS curvature of the RT surface carries information on the coefficients of the correlation function of the dual CFT$_3$. Future directions of this work considers its extension to higher dimensions and to higher-curvature gravity theories.

We have also shown that Willmore energy arises as a special case of renormalized volume formula of a two-dimensional surface. Indeed, expression \eqref{renvolume} is general, as there is no distinction between minimal and non-minimal surfaces. 
Demanding a minimal surface in a constant time slice of global AdS$_4$ bulk spacetime, makes eq.\eqref{renvolume} equivalent to the Willmore functional. The latter provides a lower bound, saturated by a circular entangling surface. This also shows that renormalized EE of a disk-like entangling region is maximal among all the shapes with the same perimeter. This is in consonance with the observations made in ref.\citep{Allais:2014ata}, which points out that the universal contribution $s_\text{univ}$ of the entanglement entropy is minimized by a circular entangling surface. At the same time, we know that $s_\text{univ}$ matches the free energy of a CFT$_3$ on a spherical $\mathbb{S}^3$ background due to the Casini-Huerta-Myers relation \citep{Casini:2011kv}. What we learnt here is that, as prescribed by eq.\eqref{Fwillmore}, $s_\text{univ}$ can be equivalently seen as the Willmore energy of $\mathbb{S}^2$.

\acknowledgments
We would like to thank I. J. Araya, C. Corral, F. D\'iaz for useful discussions and comments and P. Bueno for introducing us the concept of Willmore energy. The work of RO and GA was funded in part by FONDECYT grants No. 1170765 \textit{Boundary dynamics in anti-de Sitter gravity and gauge/gravity duality} and No. 3190314 \textit{Holographic Complexity from Anti-de Sitter gravity}. The work of JM is funded by the Comisi\'on Nacional de Investigaci\'on Cient\'ifica y Tecnol\'ogica (CONICYT) scholarship No. 21190234 and by Pontificia Universidad Cat\'olica de Valpara\'iso. The work of DRB is funded by Becas Chile (CONICYT) scholarship No. 72200301.


\appendix

\bibliography{entanglement}

\providecommand{\href}[2]{#2}\begingroup\raggedright\begin{thebibliography}{10}

\bibitem{Ryu:2006ef}
S.~Ryu and T.~Takayanagi, {\it {Aspects of Holographic Entanglement Entropy}},
  {\em JHEP} {\bf 08} (2006) 045
  [\href{http://arXiv.org/abs/hep-th/0605073}{{\tt hep-th/0605073}}].

\bibitem{Amico:2007ag}
L.~Amico, R.~Fazio, A.~Osterloh and V.~Vedral, {\it {Entanglement in many-body
  systems}},  {\em Rev. Mod. Phys.} {\bf 80} (2008) 517--576
  [\href{http://arXiv.org/abs/quant-ph/0703044}{{\tt quant-ph/0703044}}].

\bibitem{Calabrese:2009qy}
P.~Calabrese and J.~Cardy, {\it {Entanglement entropy and conformal field
  theory}},  {\em J. Phys.} {\bf A42} (2009) 504005
  [\href{http://arXiv.org/abs/0905.4013}{{\tt 0905.4013}}].

\bibitem{Casini:2009sr}
H.~Casini and M.~Huerta, {\it {Entanglement entropy in free quantum field
  theory}},  {\em J. Phys.} {\bf A42} (2009) 504007
  [\href{http://arXiv.org/abs/0905.2562}{{\tt 0905.2562}}].

\bibitem{Rangamani:2016dms}
M.~Rangamani and T.~Takayanagi, {\it {Holographic Entanglement Entropy}},  {\em
  Lect. Notes Phys.} {\bf 931} (2017) pp.1--246
  [\href{http://arXiv.org/abs/1609.01287}{{\tt 1609.01287}}].

\bibitem{Nishioka:2018khk}
T.~Nishioka, {\it {Entanglement entropy: holography and renormalization
  group}},  {\em Rev. Mod. Phys.} {\bf 90} (2018), no.~3 035007
  [\href{http://arXiv.org/abs/1801.10352}{{\tt 1801.10352}}].

\bibitem{Witten:2018lha}
E.~Witten, {\it {APS Medal for Exceptional Achievement in Research: Invited
  article on entanglement properties of quantum field theory}},  {\em Rev. Mod.
  Phys.} {\bf 90} (2018), no.~4 045003
  [\href{http://arXiv.org/abs/1803.04993}{{\tt 1803.04993}}].

\bibitem{Ryu:2006bv}
S.~Ryu and T.~Takayanagi, {\it {Holographic derivation of entanglement entropy
  from AdS/CFT}},  {\em Phys. Rev. Lett.} {\bf 96} (2006) 181602
  [\href{http://arXiv.org/abs/hep-th/0603001}{{\tt hep-th/0603001}}].

\bibitem{Lewkowycz:2013nqa}
A.~Lewkowycz and J.~Maldacena, {\it {Generalized gravitational entropy}},  {\em
  JHEP} {\bf 08} (2013) 090 [\href{http://arXiv.org/abs/1304.4926}{{\tt
  1304.4926}}].

\bibitem{VanRaamsdonk:2009ar}
M.~Van~Raamsdonk, {\it {Comments on quantum gravity and entanglement}},
  \href{http://arXiv.org/abs/0907.2939}{{\tt 0907.2939}}.

\bibitem{Maldacena:2013xja}
J.~Maldacena and L.~Susskind, {\it {Cool horizons for entangled black holes}},
  {\em Fortsch. Phys.} {\bf 61} (2013) 781--811
  [\href{http://arXiv.org/abs/1306.0533}{{\tt 1306.0533}}].

\bibitem{Arias:2019pzy}
C.~Arias, F.~Diaz and P.~Sundell, {\it {De Sitter Space and Entanglement}},
  {\em Class. Quant. Grav.} {\bf 37} (2020), no.~1 015009
  [\href{http://arXiv.org/abs/1901.04554}{{\tt 1901.04554}}].

\bibitem{Dong:2016fnf}
X.~Dong, {\it {The Gravity Dual of R\'enyi Entropy}},  {\em Nature Commun.}
  {\bf 7} (2016) 12472 [\href{http://arXiv.org/abs/1601.06788}{{\tt
  1601.06788}}].

\bibitem{Grover:2011fa}
T.~Grover, A.~M. Turner and A.~Vishwanath, {\it {Entanglement Entropy of Gapped
  Phases and Topological Order in Three dimensions}},  {\em Phys. Rev.} {\bf
  B84} (2011) 195120 [\href{http://arXiv.org/abs/1108.4038}{{\tt 1108.4038}}].

\bibitem{Liu:2012eea}
H.~Liu and M.~Mezei, {\it {A Refinement of entanglement entropy and the number
  of degrees of freedom}},  {\em JHEP} {\bf 04} (2013) 162
  [\href{http://arXiv.org/abs/1202.2070}{{\tt 1202.2070}}].

\bibitem{Calabrese:2004eu}
P.~Calabrese and J.~L. Cardy, {\it {Entanglement entropy and quantum field
  theory}},  {\em J. Stat. Mech.} {\bf 0406} (2004) P06002
  [\href{http://arXiv.org/abs/hep-th/0405152}{{\tt hep-th/0405152}}].

\bibitem{Solodukhin:2008dh}
S.~N. Solodukhin, {\it {Entanglement entropy, conformal invariance and
  extrinsic geometry}},  {\em Phys. Lett.} {\bf B665} (2008) 305--309
  [\href{http://arXiv.org/abs/0802.3117}{{\tt 0802.3117}}].

\bibitem{Casini:2011kv}
H.~Casini, M.~Huerta and R.~C. Myers, {\it {Towards a derivation of holographic
  entanglement entropy}},  {\em JHEP} {\bf 05} (2011) 036
  [\href{http://arXiv.org/abs/1102.0440}{{\tt 1102.0440}}].

\bibitem{Dowker:2010yj}
J.~S. Dowker, {\it {Entanglement entropy for odd spheres}},
  \href{http://arXiv.org/abs/1012.1548}{{\tt 1012.1548}}.

\bibitem{Klebanov:2011gs}
I.~R. Klebanov, S.~S. Pufu and B.~R. Safdi, {\it {F-Theorem without
  Supersymmetry}},  {\em JHEP} {\bf 10} (2011) 038
  [\href{http://arXiv.org/abs/1105.4598}{{\tt 1105.4598}}].

\bibitem{Jafferis:2011zi}
D.~L. Jafferis, I.~R. Klebanov, S.~S. Pufu and B.~R. Safdi, {\it {Towards the
  F-Theorem: N=2 Field Theories on the Three-Sphere}},  {\em JHEP} {\bf 06}
  (2011) 102 [\href{http://arXiv.org/abs/1103.1181}{{\tt 1103.1181}}].

\bibitem{Myers:2010xs}
R.~C. Myers and A.~Sinha, {\it {Seeing a c-theorem with holography}},  {\em
  Phys. Rev.} {\bf D82} (2010) 046006
  [\href{http://arXiv.org/abs/1006.1263}{{\tt 1006.1263}}].

\bibitem{Myers:2010tj}
R.~C. Myers and A.~Sinha, {\it {Holographic c-theorems in arbitrary
  dimensions}},  {\em JHEP} {\bf 01} (2011) 125
  [\href{http://arXiv.org/abs/1011.5819}{{\tt 1011.5819}}].

\bibitem{Casini:2012ei}
H.~Casini and M.~Huerta, {\it {On the RG running of the entanglement entropy of
  a circle}},  {\em Phys. Rev.} {\bf D85} (2012) 125016
  [\href{http://arXiv.org/abs/1202.5650}{{\tt 1202.5650}}].

\bibitem{Giombi:2014xxa}
S.~Giombi and I.~R. Klebanov, {\it {Interpolating between $a$ and $F$}},  {\em
  JHEP} {\bf 03} (2015) 117 [\href{http://arXiv.org/abs/1409.1937}{{\tt
  1409.1937}}].

\bibitem{Fei:2014yja}
L.~Fei, S.~Giombi and I.~R. Klebanov, {\it {Critical $O(N)$ models in
  $6-\epsilon$ dimensions}},  {\em Phys. Rev.} {\bf D90} (2014), no.~2 025018
  [\href{http://arXiv.org/abs/1404.1094}{{\tt 1404.1094}}].

\bibitem{Jafferis:2012iv}
D.~L. Jafferis and S.~S. Pufu, {\it {Exact results for five-dimensional
  superconformal field theories with gravity duals}},  {\em JHEP} {\bf 05}
  (2014) 032 [\href{http://arXiv.org/abs/1207.4359}{{\tt 1207.4359}}].

\bibitem{Polchinski:1983gv}
J.~Polchinski, {\it {Renormalization and Effective Lagrangians}},  {\em Nucl.
  Phys.} {\bf B231} (1984) 269--295.

\bibitem{Wilson:1973jj}
K.~G. Wilson and J.~B. Kogut, {\it {The Renormalization group and the epsilon
  expansion}},  {\em Phys. Rept.} {\bf 12} (1974) 75--199.

\bibitem{Allais:2014ata}
A.~Allais and M.~Mezei, {\it {Some results on the shape dependence of
  entanglement and R\'enyi entropies}},  {\em Phys. Rev.} {\bf D91} (2015),
  no.~4 046002 [\href{http://arXiv.org/abs/1407.7249}{{\tt 1407.7249}}].

\bibitem{Lewkowycz:2014jia}
A.~Lewkowycz and E.~Perlmutter, {\it {Universality in the geometric dependence
  of R\'enyi entropy}},  {\em JHEP} {\bf 01} (2015) 080
  [\href{http://arXiv.org/abs/1407.8171}{{\tt 1407.8171}}].

\bibitem{Fonda:2015nma}
P.~Fonda, D.~Seminara and E.~Tonni, {\it {On shape dependence of holographic
  entanglement entropy in AdS$_{4}$/CFT$_{3}$}},  {\em JHEP} {\bf 12} (2015)
  037 [\href{http://arXiv.org/abs/1510.03664}{{\tt 1510.03664}}].

\bibitem{Casini:2006hu}
H.~Casini and M.~Huerta, {\it {Universal terms for the entanglement entropy in
  2+1 dimensions}},  {\em Nucl. Phys.} {\bf B764} (2007) 183--201
  [\href{http://arXiv.org/abs/hep-th/0606256}{{\tt hep-th/0606256}}].

\bibitem{Hirata:2006jx}
T.~Hirata and T.~Takayanagi, {\it {AdS/CFT and strong subadditivity of
  entanglement entropy}},  {\em JHEP} {\bf 02} (2007) 042
  [\href{http://arXiv.org/abs/hep-th/0608213}{{\tt hep-th/0608213}}].

\bibitem{Klebanov:2012yf}
I.~R. Klebanov, T.~Nishioka, S.~S. Pufu and B.~R. Safdi, {\it {On Shape
  Dependence and RG Flow of Entanglement Entropy}},  {\em JHEP} {\bf 07} (2012)
  001 [\href{http://arXiv.org/abs/1204.4160}{{\tt 1204.4160}}].

\bibitem{Myers:2012vs}
R.~C. Myers and A.~Singh, {\it {Entanglement Entropy for Singular Surfaces}},
  {\em JHEP} {\bf 09} (2012) 013 [\href{http://arXiv.org/abs/1206.5225}{{\tt
  1206.5225}}].

\bibitem{Kallin:2014oka}
A.~B. Kallin, E.~M. Stoudenmire, P.~Fendley, R.~R.~P. Singh and R.~G. Melko,
  {\it {Corner contribution to the entanglement entropy of an O(3) quantum
  critical point in 2 + 1 dimensions}},  {\em J. Stat. Mech.} {\bf 1406} (2014)
  P06009 [\href{http://arXiv.org/abs/1401.3504}{{\tt 1401.3504}}].

\bibitem{Bueno:2019mex}
P.~Bueno, H.~Casini and W.~Witczak-Krempa, {\it {Generalizing the entanglement
  entropy of singular regions in conformal field theories}},  {\em JHEP} {\bf
  08} (2019) 069 [\href{http://arXiv.org/abs/1904.11495}{{\tt 1904.11495}}].

\bibitem{Mezei:2014zla}
M.~Mezei, {\it {Entanglement entropy across a deformed sphere}},  {\em Phys.
  Rev.} {\bf D91} (2015), no.~4 045038
  [\href{http://arXiv.org/abs/1411.7011}{{\tt 1411.7011}}].

\bibitem{Rosenhaus:2014woa}
V.~Rosenhaus and M.~Smolkin, {\it {Entanglement Entropy: A Perturbative
  Calculation}},  {\em JHEP} {\bf 12} (2014) 179
  [\href{http://arXiv.org/abs/1403.3733}{{\tt 1403.3733}}].

\bibitem{Bueno:2015lza}
P.~Bueno and R.~C. Myers, {\it {Universal entanglement for higher dimensional
  cones}},  {\em JHEP} {\bf 12} (2015) 168
  [\href{http://arXiv.org/abs/1508.00587}{{\tt 1508.00587}}].

\bibitem{Bianchi:2015liz}
L.~Bianchi, M.~Meineri, R.~C. Myers and M.~Smolkin, {\it {R\'enyi entropy and
  conformal defects}},  {\em JHEP} {\bf 07} (2016) 076
  [\href{http://arXiv.org/abs/1511.06713}{{\tt 1511.06713}}].

\bibitem{Faulkner:2015csl}
T.~Faulkner, R.~G. Leigh and O.~Parrikar, {\it {Shape Dependence of
  Entanglement Entropy in Conformal Field Theories}},  {\em JHEP} {\bf 04}
  (2016) 088 [\href{http://arXiv.org/abs/1511.05179}{{\tt 1511.05179}}].

\bibitem{Bianchi:2016xvf}
L.~Bianchi, S.~Chapman, X.~Dong, D.~A. Galante, M.~Meineri and R.~C. Myers,
  {\it {Shape dependence of holographic R\'enyi entropy in general
  dimensions}},  {\em JHEP} {\bf 11} (2016) 180
  [\href{http://arXiv.org/abs/1607.07418}{{\tt 1607.07418}}].

\bibitem{Dong:2016wcf}
X.~Dong, {\it {Shape Dependence of Holographic R\'enyi Entropy in Conformal
  Field Theories}},  {\em Phys. Rev. Lett.} {\bf 116} (2016), no.~25 251602
  [\href{http://arXiv.org/abs/1602.08493}{{\tt 1602.08493}}].

\bibitem{Ghosh:2017ygi}
A.~Ghosh and R.~Mishra, {\it {Inhomogeneous Jacobi equation for minimal
  surfaces and perturbative change in holographic entanglement entropy}},  {\em
  Phys. Rev.} {\bf D97} (2018), no.~8 086012
  [\href{http://arXiv.org/abs/1710.02088}{{\tt 1710.02088}}].

\bibitem{Carmi:2015dla}
D.~Carmi, {\it {On the Shape Dependence of Entanglement Entropy}},  {\em JHEP}
  {\bf 12} (2015) 043 [\href{http://arXiv.org/abs/1506.07528}{{\tt
  1506.07528}}].

\bibitem{Jang:2020cbm}
D.~Jang, Y.~Kim, O.-K. Kwon and D.~D. Tolla, {\it {Renormalized Holographic
  Subregion Complexity under Relevant Perturbations}},
  \href{http://arXiv.org/abs/2001.10937}{{\tt 2001.10937}}.

\bibitem{Anastasiou:2018rla}
G.~Anastasiou, I.~J. Araya and R.~Olea, {\it {Topological terms, AdS$_{2n}$
  gravity and renormalized Entanglement Entropy of holographic CFTs}},  {\em
  Phys. Rev.} {\bf D97} (2018), no.~10 106015
  [\href{http://arXiv.org/abs/1803.04990}{{\tt 1803.04990}}].

\bibitem{Babich:1992mc}
M.~Babich and A.~Bobenko, {\it {Willmore tori with umbilic lines and minimal
  surfaces in hyperbolic space}}, .

\bibitem{alexakis2010renormalized}
S.~Alexakis and R.~Mazzeo, {\it Renormalized area and properly embedded minimal
  surfaces in hyperbolic 3-manifolds},  {\em Communications in Mathematical
  Physics} {\bf 297} (2010), no.~3 621--651.

\bibitem{marques2014willmore}
F.~C. Marques and A.~Neves, {\it The willmore conjecture},  {\em Jahresbericht
  der Deutschen Mathematiker-Vereinigung} {\bf 116} (2014), no.~4 201--222.

\bibitem{willmore1996riemannian}
T.~Willmore, {\em Riemannian Geometry}.
\newblock Oxford science publications. Clarendon Press, 1996.

\bibitem{toda}
P.~Djondjorov, M.~Hadzhilazova, L.~Heller, F.~Pedit, A.~Quintino, I.~Mladenov,
  M.~Toda, V.~Vassilev and P.~Wang, {\em Willmore Energy and Willmore
  Conjecture}.
\newblock 11, 2017.

\bibitem{helfrich1973elastic}
W.~Helfrich, {\it Elastic properties of lipid bilayers: theory and possible
  experiments},  {\em Zeitschrift f{\"u}r Naturforschung C} {\bf 28} (1973),
  no.~11-12 693--703.

\bibitem{lott1988method}
N.~J. Lott and D.~Pullin, {\it Method for fairing b-spline surfaces},  {\em
  Computer-Aided Design} {\bf 20} (1988), no.~10 597--600.

\bibitem{botsch2010polygon}
M.~Botsch, L.~Kobbelt, M.~Pauly, P.~Alliez and B.~L{\'e}vy, {\em Polygon mesh
  processing}.
\newblock AK Peters/CRC Press, 2010.

\bibitem{Anastasiou:2017xjr}
G.~Anastasiou, I.~J. Araya and R.~Olea, {\it {Renormalization of Entanglement
  Entropy from topological terms}},  {\em Phys. Rev.} {\bf D97} (2018), no.~10
  106011 [\href{http://arXiv.org/abs/1712.09099}{{\tt 1712.09099}}].

\bibitem{Anastasiou:2018mfk}
G.~Anastasiou, I.~J. Araya, C.~Arias and R.~Olea, {\it {Einstein-AdS action,
  renormalized volume/area and holographic R\'enyi entropies}},  {\em JHEP}
  {\bf 08} (2018) 136 [\href{http://arXiv.org/abs/1806.10708}{{\tt
  1806.10708}}].

\bibitem{Olea:2005gb}
R.~Olea, {\it {Mass, angular momentum and thermodynamics in four-dimensional
  Kerr-AdS black holes}},  {\em JHEP} {\bf 06} (2005) 023
  [\href{http://arXiv.org/abs/hep-th/0504233}{{\tt hep-th/0504233}}].

\bibitem{Olea:2006vd}
R.~Olea, {\it {Regularization of odd-dimensional AdS gravity: Kounterterms}},
  {\em JHEP} {\bf 04} (2007) 073
  [\href{http://arXiv.org/abs/hep-th/0610230}{{\tt hep-th/0610230}}].

\bibitem{Miskovic:2014zja}
O.~Miskovic, R.~Olea and M.~Tsoukalas, {\it {Renormalized AdS action and
  Critical Gravity}},  {\em JHEP} {\bf 08} (2014) 108
  [\href{http://arXiv.org/abs/1404.5993}{{\tt 1404.5993}}].

\bibitem{Miskovic:2009bm}
O.~Miskovic and R.~Olea, {\it {Topological regularization and self-duality in
  four-dimensional anti-de Sitter gravity}},  {\em Phys. Rev.} {\bf D79} (2009)
  124020 [\href{http://arXiv.org/abs/0902.2082}{{\tt 0902.2082}}].

\bibitem{Emparan:1999pm}
R.~Emparan, C.~V. Johnson and R.~C. Myers, {\it {Surface terms as counterterms
  in the AdS / CFT correspondence}},  {\em Phys. Rev.} {\bf D60} (1999) 104001
  [\href{http://arXiv.org/abs/hep-th/9903238}{{\tt hep-th/9903238}}].

\bibitem{Kraus:1999di}
P.~Kraus, F.~Larsen and R.~Siebelink, {\it {The gravitational action in
  asymptotically AdS and flat space-times}},  {\em Nucl. Phys.} {\bf B563}
  (1999) 259--278 [\href{http://arXiv.org/abs/hep-th/9906127}{{\tt
  hep-th/9906127}}].

\bibitem{deHaro:2000vlm}
S.~de~Haro, S.~N. Solodukhin and K.~Skenderis, {\it {Holographic reconstruction
  of space-time and renormalization in the AdS / CFT correspondence}},  {\em
  Commun. Math. Phys.} {\bf 217} (2001) 595--622
  [\href{http://arXiv.org/abs/hep-th/0002230}{{\tt hep-th/0002230}}].

\bibitem{Balasubramanian:1999re}
V.~Balasubramanian and P.~Kraus, {\it {A Stress tensor for Anti-de Sitter
  gravity}},  {\em Commun. Math. Phys.} {\bf 208} (1999) 413--428
  [\href{http://arXiv.org/abs/hep-th/9902121}{{\tt hep-th/9902121}}].

\bibitem{Henningson:1998gx}
M.~Henningson and K.~Skenderis, {\it {The Holographic Weyl anomaly}},  {\em
  JHEP} {\bf 07} (1998) 023 [\href{http://arXiv.org/abs/hep-th/9806087}{{\tt
  hep-th/9806087}}].

\bibitem{Papadimitriou:2004ap}
I.~Papadimitriou and K.~Skenderis, {\it {AdS / CFT correspondence and
  geometry}},  {\em IRMA Lect. Math. Theor. Phys.} {\bf 8} (2005) 73--101
  [\href{http://arXiv.org/abs/hep-th/0404176}{{\tt hep-th/0404176}}].

\bibitem{Papadimitriou:2005ii}
I.~Papadimitriou and K.~Skenderis, {\it {Thermodynamics of asymptotically
  locally AdS spacetimes}},  {\em JHEP} {\bf 08} (2005) 004
  [\href{http://arXiv.org/abs/hep-th/0505190}{{\tt hep-th/0505190}}].

\bibitem{Anastasiou:2019ldc}
G.~Anastasiou, I.~J. Araya, A.~Guijosa and R.~Olea, {\it {Renormalized AdS
  gravity and holographic entanglement entropy of even-dimensional CFTs}},
  \href{http://arXiv.org/abs/1908.11447}{{\tt 1908.11447}}.

\bibitem{Fursaev:1995ef}
D.~V. Fursaev and S.~N. Solodukhin, {\it {On the description of the Riemannian
  geometry in the presence of conical defects}},  {\em Phys. Rev.} {\bf D52}
  (1995) 2133--2143 [\href{http://arXiv.org/abs/hep-th/9501127}{{\tt
  hep-th/9501127}}].

\bibitem{Mann:1996bi}
R.~B. Mann and S.~N. Solodukhin, {\it {Conical geometry and quantum entropy of
  a charged Kerr black hole}},  {\em Phys. Rev.} {\bf D54} (1996) 3932--3940
  [\href{http://arXiv.org/abs/hep-th/9604118}{{\tt hep-th/9604118}}].

\bibitem{Dahia:1998md}
F.~Dahia and C.~Romero, {\it {Conical space-times: A Distribution theory
  approach}},  {\em Mod. Phys. Lett.} {\bf A14} (1999) 1879--1894
  [\href{http://arXiv.org/abs/gr-qc/9801109}{{\tt gr-qc/9801109}}].

\bibitem{atiyah_lebrun_2013}
M.~Atiyah and C.~Lebrun, {\it Curvature, cones and characteristic numbers},
  {\em Mathematical Proceedings of the Cambridge Philosophical Society} {\bf
  155} (2013), no.~1 13–37.

\bibitem{Fursaev:2013fta}
D.~V. Fursaev, A.~Patrushev and S.~N. Solodukhin, {\it {Distributional Geometry
  of Squashed Cones}},  {\em Phys. Rev.} {\bf D88} (2013), no.~4 044054
  [\href{http://arXiv.org/abs/1306.4000}{{\tt 1306.4000}}].

\bibitem{Bhattacharyya:2013sia}
A.~Bhattacharyya and A.~Sinha, {\it {Entanglement entropy from the holographic
  stress tensor}},  {\em Class. Quant. Grav.} {\bf 30} (2013) 235032
  [\href{http://arXiv.org/abs/1303.1884}{{\tt 1303.1884}}].

\bibitem{Bhattacharyya:2014yga}
A.~Bhattacharyya and M.~Sharma, {\it {On entanglement entropy functionals in
  higher derivative gravity theories}},  {\em JHEP} {\bf 10} (2014) 130
  [\href{http://arXiv.org/abs/1405.3511}{{\tt 1405.3511}}].

\bibitem{Hubeny:2007xt}
V.~E. Hubeny, M.~Rangamani and T.~Takayanagi, {\it {A Covariant holographic
  entanglement entropy proposal}},  {\em JHEP} {\bf 07} (2007) 062
  [\href{http://arXiv.org/abs/0705.0016}{{\tt 0705.0016}}].

\bibitem{Hubeny:2012wa}
V.~E. Hubeny and M.~Rangamani, {\it {Causal Holographic Information}},  {\em
  JHEP} {\bf 06} (2012) 114 [\href{http://arXiv.org/abs/1204.1698}{{\tt
  1204.1698}}].

\bibitem{Bakas:2015opa}
I.~Bakas and G.~Pastras, {\it {Entanglement entropy and duality in AdS$_4$}},
  {\em Nucl. Phys.} {\bf B896} (2015) 440--469
  [\href{http://arXiv.org/abs/1503.00627}{{\tt 1503.00627}}].

\bibitem{Hubeny:2012ry}
V.~E. Hubeny, {\it {Extremal surfaces as bulk probes in AdS/CFT}},  {\em JHEP}
  {\bf 07} (2012) 093 [\href{http://arXiv.org/abs/1203.1044}{{\tt 1203.1044}}].

\bibitem{Nozaki:2013wia}
M.~Nozaki, T.~Numasawa and T.~Takayanagi, {\it {Holographic Local Quenches and
  Entanglement Density}},  {\em JHEP} {\bf 05} (2013) 080
  [\href{http://arXiv.org/abs/1302.5703}{{\tt 1302.5703}}].

\bibitem{Nozaki:2013vta}
M.~Nozaki, T.~Numasawa, A.~Prudenziati and T.~Takayanagi, {\it {Dynamics of
  Entanglement Entropy from Einstein Equation}},  {\em Phys. Rev.} {\bf D88}
  (2013), no.~2 026012 [\href{http://arXiv.org/abs/1304.7100}{{\tt
  1304.7100}}].

\bibitem{Bhattacharya:2014vja}
J.~Bhattacharya, V.~E. Hubeny, M.~Rangamani and T.~Takayanagi, {\it
  {Entanglement density and gravitational thermodynamics}},  {\em Phys. Rev.}
  {\bf D91} (2015), no.~10 106009 [\href{http://arXiv.org/abs/1412.5472}{{\tt
  1412.5472}}].

\bibitem{Witczak-Krempa:2018mqx}
W.~Witczak-Krempa, {\it {Entanglement susceptibilities and universal geometric
  entanglement entropy}},  {\em Phys. Rev.} {\bf B99} (2019), no.~7 075138
  [\href{http://arXiv.org/abs/1810.07209}{{\tt 1810.07209}}].

\bibitem{Fischetti:2016fbh}
S.~Fischetti and T.~Wiseman, {\it {A Bound on Holographic Entanglement Entropy
  from Inverse Mean Curvature Flow}},  {\em Class. Quant. Grav.} {\bf 34}
  (2017), no.~12 125005 [\href{http://arXiv.org/abs/1612.04373}{{\tt
  1612.04373}}].

\bibitem{isenberg1978science}
C.~Isenberg, {\em The science of soap films and soap bubbles}.
\newblock Tieto Cleveton, UK, 1978.

\bibitem{reilly1982mean}
R.~C. Reilly, {\it Mean curvature, the laplacian, and soap bubbles},  {\em The
  American Mathematical Monthly} {\bf 89} (1982), no.~3 180--198.

\bibitem{hopf2003differential}
H.~Hopf, {\em Differential geometry in the large: seminar lectures New York
  University 1946 and Stanford University 1956}, vol.~1000.
\newblock Springer, 2003.

\bibitem{fern2012minmax}
F.~C. Marques and A.~Neves, {\it Min-max theory and the willmore conjecture},
  2012.

\bibitem{Astaneh:2014uba}
A.~F. Astaneh, G.~Gibbons and S.~N. Solodukhin, {\it {What surface maximizes
  entanglement entropy?}},  {\em Phys. Rev.} {\bf D90} (2014), no.~8 085021
  [\href{http://arXiv.org/abs/1407.4719}{{\tt 1407.4719}}].

\end{thebibliography}\endgroup
\bibliographystyle{JHEP-2}
\label{biblio}

\end{document}